\documentclass[preprint]{JHEP3} 

\JHEPspecialurl{http://jhep.sissa.it/JOURNAL/JHEP3.tar.gz}

\usepackage{epsfig,multicol}

\def\beq{\begin{equation}}

\def\eeq{\end{equation}}

\newcommand{\bea}{\begin{eqnarray}}

\newcommand{\eea}{\end{eqnarray}}

\parskip=\itemsep               

\newcommand{\beqar}[1]{\begin{eqnarray}\label{#1}}

\newcommand{\eeqar}{\end{eqnarray}}

\def\thefootnote{\fnsymbol{footnote}} 

\title{
{\Large  \bf Inclusive Gluon Production in the QCD Reggeon Field Theory: Pomeron Loops Included. }}
\author{Tolga Altinoluk${}^{\,1}$, Alex Kovner${}^{\,1}$ and Michael Lublinsky${}^{\,2}$ 
 \\
 ${}^{\,1}$Physics Department, University of Connecticut, 2152 Hillside road, Storrs, CT 06269, USA\\
${}^{\,2}$Physics Department, State University of New York, Stony Brook, NY 11794, USA}

\abstract{We continue the study of hadronic scattering amplitudes at high energy by systematically including nonlinear effects of 
finite partonic density in hadronic wave function as well as the effects of multiple rescatterings in the scattering process.     
In this paper we derive expressions for a single inclusive gluon production amplitude and multigluon inclusive production amplitudes when the rapidities of all observed gluons are not very different. We show that at leading order these observables exhibit a semiclassical structure. Beyond the semiclassical result, we find that the gluon emission has some characteristic features different from the JIMWLK and KLWMIJ limits in that the gluons are not emitted independently in rapidity space, but have a correlated component with correlation length (in rapidity space) of order one. We demonstrate the consistency between this feature of the multigluon observables and the Hamiltonian of the QCD Reggeon Field Theory
($H_{RFT}$) derived in the companion paper \cite{rft}. We also show that the evolution of these observables with total rapidity of the process is generated by $H_{RFT}$ of \cite{rft}. We discuss whether  this evolution  is equivalent to  evolution with $H_{JIMWLK}$ as far as this set of observables is concerned.
\\

}

\begin{document}



\def\thefootnote{\arabic{footnote}} 


\section{Introduction}
This is the second of two papers devoted to the study of  hadronic observables at high energy taking into account the effects of Pomeron loops. In the first paper \cite{rft},
we derived the Hamiltonian of the QCD Reggeon Field Theory \cite{gribov} which accounts for the effects of nonlinearities in the hadronic wave function as well as multiple scattering effects in hadronic scattering, and thus fully includes  Pomeron loops. It supercedes the JIMWLK \cite{balitsky},\cite{JIMWLK},\cite{Kovchegov},\cite{cgc} and KLWMIJ \cite{klwmij} evolutions which are only applicable in parts of the parameter space. This Hamiltonian - $H_{RFT}$ governs the evolution of many hadronic observables with energy. In particular it directly gives the evolution of the forward scattering amplitude. The derivation of \cite{rft} was based on the earlier results \cite{foam} for the evolution of the hadronic wave function.

In this paper we extend our analysis to other observables. In particular we derive expressions for the amplitude of inclusive gluon production at high energy. We consider single gluon production as well as multi-gluon production when all the emitted gluons are close to each other in rapidity, so that the rapidity evolution between them does not have to be considered. There has been a lot of interest in the calculation of inclusive gluon production in the framework of the high energy evolution. Starting with \cite{kovtuch} which calculated single gluon production for DIS in the dipole model, this observable (as well as multi-gluon production) was considered beyond the dipole model limit in DIS in 
\cite{Braunincl, Baier,multigJK, diff,multig,BSV,LP} and in nucleus-nucleus 
collisions in \cite{Kovnuc,Balincl,Braunincl2,raju}. Some,  effects of the Pomeron loops which do not extend to the rapidity of the measured gluons in DIS setting were
studied in \cite{Kovpom}. 

In the present paper we extend these results by including the Pomeron loop effects using the methods of \cite{rft}. Our calculation is applicable to both the DIS situation and the nucleus-nucleus scattering. We  follow the approach and techniques developed in \cite{diff, multig}.
The extension of these results to include evolution between the observed gluons is straightforward but we leave it for future publication.

We note that very recently inclusive gluon production in nucleus-nucleus collision was discussed in \cite{raju}. Our present work has many parallels with \cite{raju} even though the methods we use are quite different.
Our results are more complete in the sense that we provide an explicit form of the observables in terms of the color charge densities of the incoming hadrons (nuclei). It is not necessary in our approach to solve dynamical equations of motion in order to find the outgoing classical color fields, as is required by the procedure of \cite{raju}. All that is necessary is to average a function of the incoming charge densities (albeit a complicated one) over the initial probability distribution. The element of averaging over the initial distributions is also present in the formalism of \cite{raju}.

We also derive a set of subleading corrections to the double inclusive and multi-gluon amplitudes, which exhibit short range rapidity correlations between the gluons in the final state. These correlations are formally subleading in $\alpha_s$ and are therefore not present in the analysis of \cite{raju}. In our approach these terms are however under control. Since our derivation of the multi-gluon observables is closely linked with the derivation of $H_{RFT}$ we are able to show quite generally that the expression for $H_{RFT}$ derived in \cite{rft} necessarily requires the presence of such short range gluon correlations in the multi-gluon observables. We stress that these are not correlations due to the evolution between the rapidity of the gluons \cite{multigJK, multig}, whose perturbative tail we also see in this calculation, but rather correlations bunching (or anti-bunching) all emitted gluons inside a rapidity interval of order unity.

Our approach allows us to show immediately that the evolution of multi-gluon observables with respect to "global" rapidity parameters is given by the RFT Hamiltonian $H_{RFT}$ derived in \cite{rft}. By "global" rapidity parameters we mean either the total rapidity of the process at fixed gluon rapidity, or the rapidity difference between the emitted gluons and the projectile; or between the emitted gluons and the target; but not the rapidity differences between the emitted gluons themselves.

Finally we discuss the question to which extent this energy evolution of the multi-gluon inclusive observables can be approximated by the JIMWLK evolution, as suggested in \cite{raju}. Formally it appears that $H_{RFT}$ can be replaced by $H_{JIMWLK}$ as long as one considers the evolution of this set of observables, with corrections to $H_{JIMWLK}$ being suppressed by powers of $\alpha_s$. However we give an example of a situation where such a formal argument fails, and thus advise caution on the subject. We believe this question merits further investigation.  

We start with recapping the notations and some formulae from \cite{rft} which we will need for our derivations. The next section does not contain any new results and is included purely in an attempt to make the present paper self-contained in terms of notations and definitions. All the formulae presented in this section are also found in \cite{rft}. Reader who is familiar with notations and results of \cite{rft} can skip directly to Section 3.
\section{High Energy Evolution as Seen From \cite{rft}.}
\subsection{Generalities}
We consider a hadronic projectile moving to the right with large energy.
We separate the degrees of freedom of the hadronic wave function into "valence" - gluons with rapidity greater than some fixed value, and "soft" - with rapidities lower than this separation rapidity. The properties of the valence component of the wave function are characterized by correlators of the color charge density operator $j^a(x)$. The wave function of the hadron can be written as 
\begin{equation}\label{psi}
|\Psi\rangle\,=\,\Omega[a,a^\dagger,j]\,|v\rangle
\end{equation}
where $a$ and $a^\dagger$ are soft gluon creation and annihilation operators.
The valence state $|v\rangle$ has no soft gluons and is therefore annihilated by the soft gluon annihilation operators
$$
a\,|v\rangle\,=\,0\,.
$$
The evolution operator $\Omega$ is a unitary operator of the Bogoliubov type
\begin{equation}\label{prod}
\Omega\,=\,{\cal C}\,{\cal B}\,.
\end{equation}
Here $\cal C$ is a coherent operator that creates the "classical" Weiszacker-Williams field
\begin{equation}\label{cohrap}
{\cal C}\,=\,\exp\left\{\,i\,\sqrt 2\,\int {d^2k\over (2\pi)^2}\ b_i^a(k)\ \int {d\eta\over 2\pi}\ 
[a^a_i(\eta,k)\,+\,a^{\dagger a}_i(\eta,-k)]\right\}\,.
\end{equation}
and $\cal B$ is a Bogoliubov type operator responsible for the leading quantum corrections.
The rapidity variable is defined as $\eta=\ln{p^-_0\over p^-}$ and the creation and annihilation operators are canonical in the rapidity basis
\begin{equation}
[a^a_i(\eta,k),a^{\dagger b}_j(\xi,p)]\,=\,(2\pi)^3\,\delta(\eta-\xi)\,\delta^2(k-p)\,.
\end{equation}

The Weiszacker-Williams field $b_i^a(x)$ 
depends only on transverse coordinates $x$ and is a two dimensional pure gauge field:
\begin{equation}
b^a_i\,=\,-{1\over g}\,f^{abc}U^{\dagger bd}\,\partial_i \,U^{dc}
\end{equation}
where $f^{abc}$ are the structure constants of the $SU(N)$ and $U^{ab}$ is an $SU(N)$ group element in the adjoint representation. The Weiszacker-Williams field is related to the valence color charge density by
\begin{equation}
\partial_i\,b^a_i(x)\,=\,j^a(x)
\end{equation}
The soft gluon operators $a(\eta)$ are labeled by rapidity variable $\eta$ as well as transverse coordinates (or momenta), color and rotational indices. They populate the rapidity interval
$Y$ equal to the parameter of the boost transformation that brings the wave function of the hadron from $\vert v\rangle$ to $\vert \Psi\rangle$. 
We do not denote this cutoff explicitly in most of our formulae and 
extend the  rapidity integration over $\eta$ to infinity for all quantities for which the integration converges. 
The $Y$ dependence is important only for divergent quantities and eventually this very dependence determines the evolution of physical observables with rapidity.

The explicit form of $\cal B$ is not important here, but its action on the gluon creation and annihilation operators is linear
\begin{equation}
\beta_\alpha\ =\ {\cal B}\, a_\alpha \, {\cal B}^\dagger\,=\,\Theta_{\alpha\beta}[j]\ a_\beta\ +\ \Phi_{\alpha\beta}[j]\ a^\dagger_\beta\,,\ \ 
\ \ \ \ \ \ \ \ \ \ \ \beta^\dagger_\alpha\ ={\cal B}\, a^\dagger_\alpha \, {\cal B}^\dagger\ =\ \Theta^*_{\alpha\beta}[j]\ a^\dagger_\beta\ +\ \Phi^*_{\alpha\beta}[j]a_\beta\,
\label{ML}
\end{equation}
It turns out to be useful to define separately the matrix
\begin{equation}
N(p,k)\,=\,\Theta(p,k)-\Phi(p, -k)\,, 
\end{equation}
where we have indicated explicitly only the transverse momentum dependences.
The transformation matrices $\Theta$, $\Phi$ and $N$ depend on the valence color charge density and where calculated in \cite{rft}. The explicit form of $\Theta$ and $\Phi$ is not important for us in this paper. The expression for $N$ is
\begin{equation}
N(p,\eta;k,\xi)= {2i\over e^{-(\xi-\eta)} -e^{\xi-\eta}-i\epsilon}
\langle p\vert (1-2l)\vert k\rangle-\langle p\vert (1-2L){2i\over {D^2\over\partial^2}e^{-(\xi-\eta)}-{\partial^2\over D^2}e^{\xi-\eta}+i\epsilon}\vert k\rangle
\end{equation}
where $D$ is the transverse covariant derivative in the background filed $b$, and the longitudinal projectors $l$ and $L$ are defined as
\begin{equation}\label{projectors}
l^{ab}_{ij}\equiv {\partial_i\partial_j\over \partial^2}\delta^{ab};\ \ \ \ \ \ \ \ \ \ \ \ \ \ 
L^{ab}_{ij}\equiv D_i^{ae}(b)\left[{1\over D^2(b)}\right]^{ed}D_j^{db}(b)
\end{equation}

The vacuum of the operator $\beta$ as calculated in \cite{rft} is
\begin{equation}\label{betavac}
\vert 0\rangle_\beta={\cal B}\vert 0\rangle=\ e^{\,{1\over 4}\, {\rm Tr}\, \ln(1\,-\,\Lambda^\dagger\,\Lambda)}\ e^{\,-{1\over 2}\,a^\dagger\,
\Lambda \,a^\dagger}\ |0\rangle
\end{equation}
with
\begin{equation}\label{Lambda}
\Lambda\ =\ \Theta^{-1}\,\Phi\,.\
\end{equation}

Eqs.(\ref{psi},\ref{cohrap},\ref{betavac}) characterize the wave function of a hadron with rapidity $Y$.
This hadron, which we call the "projectile" scatters on another hadron- the "target". The target is specified by a distribution of color fields $\alpha^a$. Calculation of many interesting observables in the scattering process, including multi-gluon inclusive amplitudes, can be represented as the double average over the projectile and the target wave functions of some operator $\hat O$
\begin{equation}\label{Ot}
\langle\,\bar O\,\rangle\,_T\,=\,\int D\alpha\,W^T[\alpha]\,\,\bar O\,; \, \ \ \ \ \ \ 
\  \ \ \ \ \ \ \ \ \bar O\,\equiv\,\langle v\vert\hat O[j]\vert v\rangle\,=\,\int Dj\,W^P[j]\,O[j]\,.
\end{equation}
Here $W^P[j]$ is a weight functionals representing the probability
distributions of the projectile color charge $j$, while $W^T[\alpha]$ is the same for the target fields $\alpha$. 
The rapidity evolution of this observable is given by $H_{RFT}$ (\ref{hrft})
in the following sense \cite{JIMWLK,rft}
\begin{equation}
{d\over dY}\bar O\  =\ -
\int Dj\  W^P[j]\ H_{RFT}\left[j,{\delta\over \delta j}\right]\ O[j]\,.
\end{equation}
or equivalently
\begin{equation}
{d\over dY}\,W_Y^P[j]\ =\ -\,H_{RFT}\left[j,{\delta\over \delta j}\right]\ W_Y^P[j]\,.
\end{equation}
The Hamiltonian $H_{RFT}$ is formally defined as the expectation value of the unitary operator $\hat R_a$ in the hadronic wave function $\vert \Psi\rangle$
\begin{equation}\label{defin}
H_{RFT}\left[j,{\delta\over \delta j}\right]\,=\,-{d\over dY}\langle 0|\Omega^\dagger[j,a,a^\dagger] \,
\hat R_a\, \Omega[j,a,a^\dagger]|0\rangle|_{Y=0}\,.
\end{equation}
with
\begin{equation}\label{R}
\hat R_a\,\equiv\, e^{\ \int_x j^a_{soft}(x)\ {\delta\over \delta j^a(x)}}
\end{equation}
and
\begin{equation}
j^a_{soft}(x)\,=\,g\,\int {d\eta\over 2\pi}\, a^{\dagger\,b}(\eta,x) \ T^a_{bc}\ a^c(\eta,x)
\end{equation}
The operator $\hat R_a$ shifts the valence color charge density by the charge density of the soft gluons. This reflects the fact that in the boosted wave function not only valence, but also soft gluons participate in the scattering, and thus the effective color charge density  
must include both $j$ and $j_{soft}$.

$H_{RFT}$ was calculated in \cite{rft}. To summarize the results of \cite{rft} we have to introduce (alas - at the risk of being repetitive) some additional notations.

\subsection{Definitions}
First, since the color charge density operators do not commute, we need to keep track of the ordering in eq.(\ref{defin}). As a result we have to define the right and left color charge density operators $J_R$ and $J_L$. The labels refer to their order in the calculation of the matrix element of $O$ and the Hamiltonian $H_{RFT}$.
The operator $\Omega$ in eq.(\ref{defin}) has to be understood as a function of the right charge operator $J_R$, while the operator $\Omega^\dagger$ is a function of $J_L$. 
This is important, since observable $O$ generically depends on the color charge density, and in the calculation of its matrix element in $\vert\Psi\rangle$ it always appears sandwiched between $\Omega^\dagger$ and $\Omega$. Defining $J_R$ and $J_L$ specifies the relative ordering between the color charge operators in the wave function and those in the observable. 
As shown in \cite{kl4}, the right and left charge densities are  related by the 
rotation with unitary matrix $R$ (here $t$ is the ordering variable \cite{klwmij})
\begin{equation}
R^{ab}(x)\ =\ \left[{\cal P}\,\exp\{g\int_0^1 dt\  T^c\ {\delta\over\delta j^c(x,t)}\right]^{ab}\,.
\end{equation}
The matrix $R$ has the meaning of a scattering matrix of the gluon of the target on the projectile\cite{kl4}. As discussed, for example in \cite{yinyang}, it can be taken as the basic degree of freedom of the Reggeon Field Theory. We can then write \cite{kl4}
\begin{eqnarray}
&&gJ_R^a(x)=-g{\rm tr} \left\{R(x)T^{a}{\delta\over \delta R^\dagger(x)}\right\};\\
&&gJ_L^a(x)=
-g{\rm tr} \left\{T^{a}R(x){\delta\over \delta R^\dagger(x)}\right\};\nonumber\\
&&J_L^a(x)\,\,=\,\,[R(x)\,J_R(x)]^a\,.\nonumber
\end{eqnarray}
Accordingly we define two classical fields, $b_R$ and $b_L$:
\begin{eqnarray}
b^a_{Ri}\,&=&\,-{1\over g}\,f^{abc}U^{\dagger bd}[J_R]\,\partial_i \,U^{dc}[J_R]\equiv\,-{1\over g}\,f^{abc}U_R^{\dagger bd}\,\partial_i \,U_R^{dc}; \nonumber\\
b^a_{Li}\,&=&\,-{1\over g}\,f^{abc}U^{\dagger bd}[J_L]\,\partial_i \,U^{dc}[J_L]\equiv\,-{1\over g}\,f^{abc}U_L^{\dagger bd}\,\partial_i \,U_L^{dc}
\end{eqnarray}
The matrices $\Theta$, $\Phi$ and $\Lambda$ also "bifurcate" into Right and Left quantities. 
\begin{equation}
\Lambda_R\equiv
\Lambda[J_R]\,=\,\Theta^{-1}[J_R]\,\Phi[J_R]; \ \ \ \  \ \ \ \ \ \ \ \ \ \ \ \ \ \ \ \ \ \ \ \Lambda_L\equiv
\Lambda[J_L]\,=\,\Theta^{-1}[J_L]\,\Phi[J_L]\,.
\end{equation}

It is thus clear that for the purpose of the calculation of the average of any observable the proper ordering of the factors of the charge density is equivalent to the substitution
\begin{equation}
\Omega^\dagger\,\rightarrow\, \Omega^\dagger_L\,=\,{\cal B}^\dagger_L\,{\cal C}^\dagger_L; 
\ \ \ \ \ \ \ \ \ \ \ \ \ \Omega\,\rightarrow\, \Omega_R\,=\,{\cal C}_R\,{\cal B}_R
\end{equation}
where the subscript $L$ ($R$) indicates that the respective operator depends on $J_L$ ($J_R$). 
Due to the presence of the operator $\hat R_a$ in eq.(\ref{defin}) (and in other similar averages) it turns out that the Left quantities are always rotated by the matrix $R$.
It is thus convenient to introduce the barred quantities
\begin{equation}
\bar \Omega_L\,\equiv\, \bar {\cal C}_L\,\bar {\cal B}_L;\ \ \ \ \ \ \ \ \ \ \ \ \ \ \ \ 
\bar {\cal B}_L\,\equiv\,\hat R^\dagger_a \,{\cal B}_L\,,\ \ \ \ \ \ \ \ \ \ \ \ \ \ \ \ 
\bar {\cal C}_L\,\equiv\, \hat R^\dagger_a\, {\cal C}_L\, \hat R_a\,.
\end{equation}
and accordingly
\begin{equation}
\bar b_{L\,i}^{a}\,\equiv\,R^{\dagger ab}\,b_{L\,i}^{b}; 
\ \ \ \ \ \ \ \ \ \ \ \ \ \ \ \ \ \ \  \bar\Lambda_L\,\equiv\,R^\dagger\,\Lambda_L\,R;\, \ \ \ \ \ \ \ \ \bar N_L=N_LR
\end{equation}
Finally we introduce another set of matrices
\begin{eqnarray}\label{ke1}
&&K_{\alpha\beta}\,=\,(\Theta\,\bar \Theta^\dagger\,-\,\Phi\,\bar \Phi^\dagger)_{\alpha\beta};\, \ \ \ \ \ \ \ \ \ \ \ \ 
E_{\alpha\beta}\,=\,(\Phi\,\bar \Theta^T\,-\,\Theta\,\bar \Phi^T)_{\alpha\beta}\,; \\
\label{defXi}&&\Xi\ =\ K^{-1}\,E\,.
\end{eqnarray}
which satisfy
\begin{eqnarray}\label{identi}
&&KE^T-EK^T=0,\ \ \ \ \ \ KK^\dagger-EE^\dagger=1;  \ \ \ K^\dagger K-E^TE^*=1; \ \ \ \ \ 
K^\dagger E-E^TK^*=0\,; \\
\label{kk}&&{1\over 1-\Xi\,\Xi^\dagger}\ =\ K^\dagger\, K\ ;\ \ \ \ \ \ \ \  \ \ \ \ \ \ \ \ 
{1\over 1-\Xi^\dagger\,\Xi}\ =\ K^T\, K^*\,.
\end{eqnarray}
Above we have omitted the indices $L$ and $R$. All the barred quantities depend on $J_L$ while the unbarred ones - on $J_R$. We will stick to this practice in the following, whenever it does not lead to confusion.

The matrices $K$, $E$ and $\Xi$ are the analogs of the Bogoliubov transform eq.(\ref{ML}) for transformation between the operators $\beta_R$ ( defined by eq.(\ref{ML}) with $\Theta_R$, etc.) and $\bar\beta_L$ ( defined by eq.(\ref{ML}) with $\bar \Theta_R$, etc.).
The matrices $K$ and $E$, which we will need in the following have been calculated in \cite{rft}

\begin{eqnarray}\label{K}
&&K(p,\eta;k,\xi)\,=\,i\langle p\vert 2(1-l-L_R)R^\dagger(l- L_L)+(1-l-L_R)(\bar\Delta-\Delta^\dagger_R)R^\dagger(1-l- L_L)\vert k\rangle\nonumber\\
&&+{i\over 1-e^{\xi-\eta}+i\epsilon}\langle p \vert(1-2l)R^\dagger(1-2l)\vert k\rangle-i\langle p\vert (1-2L_R){1\over 1-{D^2_R\over \bar D^2}e^{\xi-\eta}-i\epsilon}R^\dagger(1-2L_L)\vert k\rangle\nonumber\\ && \nonumber \\
&&E(p,\eta;-k,\xi)\,=\,i\langle p\vert 2(1-l-L_R)R^\dagger(l- L_L)+(1-l-L_R)(\bar\Delta -\Delta^\dagger_R)R^\dagger(1-l- L_L)\vert k\rangle
\nonumber \\
&&+{i\over 1+e^{\xi-\eta}}\langle p \vert(1-2l)R^\dagger(1-2l)\vert k\rangle-i\langle p\vert (1-2L_R){1\over 1+{D^2_R\over \bar D^2}e^{\xi-\eta}}R^\dagger(1-2L_L)\vert k\rangle
\end{eqnarray}
with
\begin{equation}
\bar D\ \equiv\  R^\dagger\, D_L\,R ; \ \ \ \ \ \ \ \ \ \ \ \ \ \ \ \bar\Delta\ =\ R^\dagger\,\Delta_L\,R\,.
\end{equation}
In eq. (\ref{K}) $\Delta$ is an operator in the transverse space, which we do not specify, since it does not appear in the following, $D_R$ and $D_L$ are transverse covariant derivatives in the background fields $b_R$ and $b_L$ respectively, and longitudinal projectors $L_{R(L)}$ are defined as in eq.(\ref{projectors}).
In eq.(\ref{K}) all $D_R$ are ordered to the left of $D_L$.
 
\subsection{$H_{RFT}$}

The result of \cite{rft} for $H_{RFT}$ to leading order in $\alpha_s$ 
in the concise matrix  notation  is
\begin{eqnarray}\label{hrft}
H_{RFT}={d\over dY}\Bigg\{\,&&\int_{x,y,z,u}\left[b(x)-\bar b(x)\right]\int_{\eta,\xi,\lambda\zeta} 
\bar N^\dagger(\eta,x;\xi,y)\,K^{-1}(\xi,y;\lambda, z)\,N(\lambda,z;\zeta, u)\,\left[b(u)-\bar b(u)\right]\nonumber \\
&+&{1\over 4}{\rm Tr}\  \ln\ ( KK^\dagger)\Bigg\}|_{Y=0}\,.
\end{eqnarray}
The ${\rm Tr \ ln}$ term in this expression is subleading in $\alpha_s$ and will not be considered any further. While $b$ is independent of rapidity,
both $N$ and $K$ depend on rapidity differences only. Introducing
\beq
N_\perp(x,y)\,\equiv \,\int_{\eta-\xi} N(x,\eta,y,\xi)\,=  \ [1-l-L]\,
\eeq
which will reappear in the next section
and performing the rapidity integrations the result for the first term in eq.(\ref{hrft}) can be recast in the form
\begin{eqnarray}\label{hg}
H_{RFT}& =&
{1\over \pi}\,[b_RR^\dagger- b_L\,]  \,\left(1-l- L_L\right) \nonumber \\ && \times\  
\left[(1-2l)\,R^\dagger\,(1-2l)\,+\,(1-2L_R)\,R^\dagger\,(1-2L_L)\right]^{-1}\ \left(1-l- L_R\right)\ [b_R-R^\dagger \,b_L]\,.
\end{eqnarray}
Reinstating explicitly transverse coordinate dependences this reads 
\begin{eqnarray}\label{hg1}
H_{RFT}&=&
{1\over 8\pi^3 }\,\int_{x,y,z,\bar z}[b_{Ri}^b(x)\,R^{\dagger ba}(x)\,-\,b^a_{Li}(x)]\ \left[\delta_{ij}{1\over (x-z)^2}\,-\,
2\,{(x-z)_i\,(x-z)_j\over (x-z)^4}\right]\nonumber \\ \nonumber \\ 
&\times&\left[\delta^{ac}\,+\, [U_L^\dagger(x)\,U_L(z)]^{ac}\right]
\  \tilde K^{-1\,cd}_{\perp jk}(z,\bar z)\ \left[\delta_{kl}{1\over (y-\bar z)^2}\,-\,2\,{(y-\bar z)_k\,(y-\bar z)_l \over (y-\bar z)^4}\right]
\nonumber \\ \nonumber \\ 
&\times&\left[\delta^{de}\,+\, [U_R^\dagger(\bar z)\,U_R(y)]^{de}\right]\ [b_{Rl}^e(y)\,-\,R^{\dagger ef}(y)\,b^f_{Lk}(y)]
\end{eqnarray}
with
\begin{eqnarray}
\tilde K^{\ ab}_{\perp ij}(x,y)&=&{1\over 2\,\pi^2}\,\int_z\left[\delta_{ik}{1\over (x- z)^2}\,-\,2\,{(x- z)_i\,(x- z)_k\over (x- z)^4}\right]
\ \left[\delta_{kj}{1\over (z-y)^2}\,-\,2\,{(z-y)_k\,(z-y)_j\over (z-y)^4}\right]\nonumber\\
&\times&\left\{R^{\dagger ab}(z)\,+\,\left[U_R^\dagger(x)\,U_R(z)\,R^{\dagger }(z)\,U^\dagger_L(z)\,U_L(y)\right]^{ab}\right\}
\end{eqnarray}

With all the definitions in place we can now proceed to calculation of the multi-gluon spectrum.

\section {The amplitudes $Q$ and the scattering amplitude.}
We will start by rederiving the results of \cite{rft} in a slightly more general way which later will  allow us to calculate additional observables. First, we remind the reader the general formalism of \cite{yinyang}, which we will use here.  The matrix element eq.(\ref{defin}) pertinent to the calculation of $H_{RFT}$ can be represented, using the optical theorem as
\begin{equation}
\langle 0\vert 1-\Omega^\dagger\,\hat R_a\,\Omega\vert0\rangle\,
=\,{1\over 2}\langle 0\vert (1-\Omega^\dagger\,\hat R_a^\dagger\,\Omega)\
(1-\Omega^\dagger\,\hat R_a\,\Omega)\vert0\rangle\,.
\end{equation}
 This assumes that the amplitude is real, which in our case can be verified directly by examining eq.(\ref{hg}).
 Introducing the complete basis of intermediate states and defining the amplitudes
 \begin{equation}\label{Q}
 Q_n(x_i,\eta_i)\,=\,\langle x_1,\eta_1;...;x_n,\eta_n\vert (1-\Omega^\dagger\,\hat R_A\,\Omega)\vert 0\rangle
\,=\, \langle n|\,\bar\Omega^\dagger\,\Omega\vert 0\rangle
 \end{equation}
 we can write
 \begin{equation}\label{qsquare}
 H_{RFT}\ =\ {1\over 2}\,{d\over d Y}\sum_{n=1}^\infty \,\int\,\Pi_{i=1}^n\,[ dx_id\eta_i]\ 
 Q^\dagger_n(x_i,\eta_i)\,Q_n(x_i\eta_i)\,|_{Y=0}\,.
 \end{equation}
 As explained in \cite{yinyang}, the amplitudes $Q_n$ has the meaning of emission amplitudes of $n$ gluons in one step of evolution. The knowledge of $Q_n$ in principle allows one to construct also exclusive observables, for example constraining the number of particles produced in the final state. In the next section we will use $Q_n$ to calculate single and double inclusive gluon production.
 
 Our aim is to write the expression for $Q_n$
in the normal ordered form, so that the amplitudes  can be easily read off.
 \begin{equation} 
 \bar\Omega^\dagger\,\Omega\,\vert 0\rangle\,=\,\bar {\cal B}^\dagger\,\bar {\cal C}^\dagger\,
{\cal C}\,{\cal B}\,\vert 0\rangle\,=\,\bar {\cal B}^\dagger\
 e^{i\sqrt 2(\bar b-b)\,(a+a^\dagger)}\ \bar {\cal B}\,\bar {\cal B}^\dagger\, {\cal B}\,\vert 0\rangle\,.
 \end{equation}
 where 
 \begin{equation}
 \sqrt 2(\bar b-b)\,(a+a^\dagger)\,\equiv \,\sqrt 2\int d^2x\Big[\bar b^a_i(x)-b^a_i(x)\Big]\int {d\eta\over 2\pi}\Big[a^a_i(x,\eta)+a^{a\dagger}_i(x,\eta)\Big]\,.
 \end{equation}
 For the first factor, using the fact that $\bar {\cal B}$ is the Bogoliubov operator with the action similar to eq.(\ref{ML}) with $\Theta,\ \Phi\rightarrow \bar\Theta,\ \bar\Phi$ we have
 \begin{equation} 
 \bar {\cal B}^\dagger \,e^{i\sqrt 2\,(\bar b-b)\,(a+a^\dagger)}\,\bar {\cal B}\,=\,
e^{i\sqrt 2\,(\bar b-b)\,(\bar N^\dagger \,a+\,\bar N^T\,a^\dagger)}\,=\,e^{-(\bar b-b)\,\bar N^\dagger\,\bar N\,(\bar b-b)}
\,e^{i\sqrt 2\,(\bar b-b)\,\bar N^T\,a^\dagger}\,e^{i\sqrt 2\,(\bar b-b)\,\bar N^\dagger a}\,.
 \end{equation}
The state $\bar {\cal B}^\dagger\, {\cal B}\,\vert 0\rangle$ is the vacuum of the operator
\begin{equation}\label{KE}
\tilde \beta\,=\,\bar {\cal B}^\dagger \,{\cal B}\, a\,\bar {\cal B}\,{\cal B}^\dagger\,=\,K\,a\,+\,E\,a^\dagger;
\end{equation}
with $K$ and $E$ defined in eq.(\ref{ke1}). Comparing eq. ({\ref{KE}) with eq. (\ref{ML}), 
we can use eq. (\ref{betavac}) with the substitution $\Lambda\rightarrow \Xi$
to write
\begin{equation}
 \bar {\cal B}^\dagger\, {\cal B}\,\vert0\rangle\,=\,
e^{{1\over 4}\,Tr\ln(1-\Xi^\dagger\Xi)}\,e^{-{1\over 2}\,a^\dagger\,\Xi\,
\,a^\dagger}\vert 0\rangle\,.
 \end{equation}
 Now using the fact that the annihilation operator can be represented as $a={\delta\over \delta a^\dagger}$ we can write
 \begin{equation}
e^{i\sqrt 2\,(\bar b-b)\,\bar N^\dagger\, a} \, e^{-{1\over 2}\,a^\dagger\,\Xi\, a^\dagger}\vert 0\rangle
\,=\,e^{-{1\over 2}\,\left(a^\dagger+i\sqrt 2\,(\bar b-b)\,\bar N^\dagger\right)\  \Xi \ 
\left( a^\dagger+i\sqrt 2\,(\bar b-b)\,\bar N^\dagger\right)}\vert 0\rangle\,.
 \end{equation}
 And so, finally the normal ordered form is 
 \begin{equation}\label{nor}
 \bar\Omega^\dagger\,\Omega\vert \,0\rangle\,=\,e^{{1\over 4}\,Tr\ln(1-\Xi^\dagger\Xi)}\,
e^{-(\bar b-b)\,\left[\bar N^\dagger\bar N-\bar N^\dagger\,\Xi\, \bar N^*\right]\,(\bar b-b)}\ 
e^{i\sqrt 2\,(\bar b-b)\,\left[\bar N^T-\bar N^\dagger\,\Xi\right]\,a^\dagger}\ 
e^{-{1\over 2}\,a^\dagger\,\Xi \, a^\dagger}\,|0\rangle\,.
 \end{equation}
 Using the relations eqs.(\ref{identi}, \ref{defXi}) and the definition of $\Xi$ in terms of $\Theta,\Phi,\bar\Theta,\bar \Phi$ it is straightforward to show
 \begin{equation}\label{iden2}
 N=K\bar N-E\bar N^*  ; \ \ \ \ \ \ \ \ \ \ \ \ \ \ \bar N=K^\dagger N+E^TN^*\,.
 \end{equation}
  Thus we can rewrite eq.(\ref{nor}) as  
 \begin{equation}\label{nor1}
 \bar\Omega^\dagger\,\Omega\,\vert 0\rangle\,=\,
e^{-{1\over 4}\,Tr\ln(K^\dagger K)}\ e^{\,-\,(\bar b-b)\,\bar N^\dagger \,K^{-1}\,N\,(\bar b-b)}\ 
e^{\,i\,\sqrt 2\,a^\dagger\, K^{-1}\,N\,(\bar b-b)}\ 
e^{\,-\,{1\over 2}\,a^\dagger\,\Xi\, a^\dagger}\,\vert 0\rangle\,.
 \end{equation}
Taking the overlap of eq.(\ref{nor1}) with the vacuum ($\langle 0|\,\bar\Omega^\dagger\Omega\,\vert 0\rangle$)
 we  reproduce the result of \cite{rft} eq.(\ref{hrft}), 
while the last two factors in eq.(\ref{nor1}) allow us to calculate $Q_n$ in a straightforward manner.
 
 \subsection*{Scattering amplitude once again.}
 
 Before moving on to discussion of inclusive gluon spectrum, we rederive the expression for the scattering amplitude and for the RFT Hamiltonian using eq.(\ref{qsquare}). This will explicitly establish the equivalence of the two approaches and will also serve as a consistency check on our expressions for the amplitudes $Q_n$. According to eq.(\ref{qsquare}) we need to extract the term linear in $Y$ from the sum of squares of the amplitudes $Q_n$ integrated over the transverse coordinates and rapidities. The amplitudes $Q_n$ are obtained by expanding the exponent in eq.(\ref{nor1}) to appropriate order. In general therefore the calculation will involve expressions of the type
 \begin{equation}\label{expa}
 \langle 0\vert\,[(\bar b-b)\,N^\dagger\, K^{\dagger -1}\,a]^m\ 
[a\,\Xi^\dagger \,a]^k\ [a^\dagger\, K^{-1}\,N\,(\bar b-b)]^l\ [a^\dagger\,\Xi\, a^\dagger]^n\,\vert 0\rangle\,.
 \end{equation}
Here $m,\,n,\,l,\,k$ are some integer powers to be summed over.
 For the purpose of the present discussion we will think of the matrix $\Xi(\xi-\eta)$ as short range in rapidity space. Although this is not really the case, as discussed in detail in  \cite{rft}, the nonlocal in rapidity terms have to be subtracted from the final result, since they lead to higher powers of $Y$.  Recall that the field $\bar b-b$ does not depend on rapidity. Thus after contracting all $a$'s with $a^\dagger$'s in eq.(\ref{expa}) we will generate expressions of two types and their products (for simplicity of notation we drop in this schematic discussion the factors of $N\,K^{-1}$ which accompany powers of $(\bar b-b)$)
 \begin{equation}
 (\bar b-b)\,\int_{\eta\, \xi}\,[\Xi^\dagger\, \Xi]^k(\eta,\xi)\,(\bar b-b) ;
\ \ \ \ \ \ \ \ \ \ \  \ \ \ \ \ \ Tr[\Xi^\dagger\,\Xi]^l\,.
 \end{equation}
 It is easy to see that each one of these terms is of order $Y$. Thus we only need to consider contractions in eq.(\ref{expa}) that lead to appearance of only one such "connected" term, rather than product of two or more - the "disconnected" graphs . This has an immediate consequence that only terms of order $(\bar b-b)^2$ have to be kept. Higher powers of $b$ necessarily lead to disconnected graphs and thus necessarily to higher powers of $Y$. On the other hand, the terms which do not contain any powers of $b$ 
are suppressed by a power of coupling constant $\alpha_s$. Thus we are lead to the conclusion that the only terms that contribute to $H_{RFT}$ are connected contractions of order $(\bar b-b)^2$. 
With this in mind the calculation becomes straightforward. 

Let us consider odd and even $n$'s separately.
\begin{equation}
Q_{2n+1}\,=\,i\,(-1)^n\,{\sqrt 2\over 2^n n!}\,\langle n\vert \,[a^\dagger\, K^{-1}\,N\,(\bar b-b)]\ 
[a^\dagger\,\Xi \, a^\dagger]^n\,\vert 0\rangle \,.
\end{equation}
Then
\begin{equation}\label{nodd}
\int dx_i\,d\eta_i\, Q^\dagger_{2n+1}(x,\eta)\,Q_{2n+1}(x,\eta)\,=\,2\,(\bar b-b)\,N^\dagger\, K^{\dagger-1}\,
[\Xi\,\Xi^\dagger]^n\,K^{-1}\,N\,(\bar b-b)\,.
\end{equation}
Thus
\begin{equation}
\sum_{n=0}^\infty Q^\dagger_{2n+1}\,Q_{2n+1}\,=\,2\,(\bar b-b)\,N^\dagger\, K^{\dagger-1}\,
{1\over 1-\Xi\,\Xi^\dagger}\,K^{-1}\,N\,(\bar b-b)\,=\,2\,(\bar b-b)\,N^\dagger\, N\,(\bar b-b)
\end{equation}
where we have used eq.(\ref{kk}).
For even $n$ we have
\begin{equation}
Q_{2n}\,=\,(-1)^n\,{1\over 2^nn!}\,\langle n\vert\, 2\,n\,[a^\dagger\, K^{-1}\,N\,(\bar b-b)]^2\,
[a^\dagger\,\Xi \, a^\dagger]^{n-1}\,+\,
[a^\dagger\,\Xi \, a^\dagger]^n\,\vert 0\rangle\,.
\end{equation}
So
\begin{equation}\label{neven}
\int dx_i\,d\eta_i\,Q^\dagger_{2n}(x,\eta)\,Q_{2n}(x,\eta)\,=\,\left[(\bar b-b)\,N^T\, K^{-1\,T}\,
\Xi^\dagger\,[\Xi\,\Xi^\dagger]^{n-1}\,K^{-1}\,N\,(\bar b-b)\,+\,h.c.\right]\,.
\end{equation}
Thus
\begin{eqnarray}
\sum_{n=1}^\infty&& Q^\dagger_{2n}\,Q_{2n}\ = \nonumber \\
&&=\,\left[(\bar b-b)\,N^T \,K^{-1\,T}\,\Xi^\dagger\,
{1\over 1-\Xi\,\Xi^\dagger}\,K^{-1}\,N\,(\bar b-b)\,+\,(\bar b-b)\,N^\dagger \,K^{-1\,\dagger}\,
{1\over 1-\Xi\,\Xi^\dagger}\,\Xi\, K^{-1*}\,N^*\,(\bar b-b)\right]\nonumber\\ \nonumber \\
&&=\,\left[(\bar b-b)\,N^T \,K^{-1\,T}\,E^\dagger\, N\,(\bar b-b)\,+\,(\bar b-b)\,
N^\dagger \,E\, K^{-1*}\,N^*\,(\bar b-b)\right]\,.
\end{eqnarray}
Altogether
\begin{equation}
\sum_{n=1}^\infty Q^\dagger_n\,Q_n\,=\,(\bar b-b)\,\left[2\,N^\dagger\, N\,+\,N^T\, K^{-1\,T}\,E^\dagger\, N\,+\,
N^\dagger\, E\, K^{-1*}\,N^*\right]\,(\bar b-b)\,.
\end{equation}
Finally using eq.(\ref{iden2}) it can be recast in the form
\begin{equation}
\sum_{n=1}^\infty Q^\dagger_n\,Q_n\,=\,(\bar b-b)\,\left[\bar N^\dagger \,K^{-1}\, N\,+\,h.c.\right]\,
(\bar b-b)\,=\,2\,(\bar b-b)\,\bar N^\dagger\, K^{-1}\, N\,(\bar b-b)
\end{equation}
which reproduces $H_{RFT}$ - the first term in eq.(\ref{hrft}).

We note without proof, that were we to keep also subleading in $\alpha_s$ terms in $Q^\dagger_{2n}Q_{2n}$ we would also reproduce the second term in eq.(\ref{hrft}). 
\section{Inclusive gluon production.}
\subsection{Single gluon inclusive production.}

We now turn to derivation of the amplitude $dn(\eta,k)/ d\eta$ for inclusively produce
gluon with rapidity $\eta$ and transverse momentum $k$.
Here we follow the approach of \cite{diff, multig}.

 The single gluon inclusive amplitude is defined as 
\beq\label{gav}
{dn(\eta,k)\over d\eta}\,=\,\int Dj\,DS\,W^P[j]\,  O_g(k,\eta)\, W^T[S]
\eeq
with the gluon operator to be measured in the collision 
\begin{eqnarray}\label{si}
 O_g(k,\eta)&=&{1\over 2\pi}\, \langle 0\vert\,\Omega^\dagger\, \hat S^\dagger\, \Omega\ 
 a^{\dagger a}_i(\eta,k)\,a^{a}_i(\eta, k)\ \Omega^\dagger\, \hat S \,\Omega\,\vert 0\rangle\ =\nonumber \\
&=&{1\over 2\pi}\, \int_{x,y} e^{i\,k\,(x-y)}\ \langle 0\vert\,\Omega^\dagger\, \hat S^\dagger\, \Omega\ 
 a^{\dagger a}_i(\eta,x)\,a^{a}_i(\eta, y)\ \Omega^\dagger\, \hat S \,\Omega\,\vert 0\rangle\,.
\end{eqnarray}
The state $\vert 0\rangle$ as before is the vacuum of the soft gluon Hilbert space.
The operator $\hat S$ here is the second quantized $S$-matrix operator, which acts as a color rotation on the gluon creation and annihilation operators as well as on the color charge density in the operator $\Omega^\dagger$
\begin{equation}
\hat S^\dagger\, a^a(x)\,\hat S\,=\,S^{ab}(x)\,a^b(x); \ \ \ \ \ \ \ \ \ \ \ \ \hat S^\dagger\, j^a(x)\,\hat S\,=
\,S^{ab}(x)\,j^b(x)\,.
\end{equation}
The unitary matrix $S(x)$ is the scattering matrix of a single gluon in the color field of the target. The target average in eq.(\ref{si}) amounts to integrating over $S(x)$ with a weight function determined by the wave function of the target\cite{diff},\cite{multig}.  

The definition eq.(\ref{si}) involves objects almost identical to the amplitudes $Q_n$ discussed above.
In fact the operator $\bar\Omega^\dagger\,\Omega$  becomes identical to the operator
 $\Omega^\dagger\hat S\Omega$ upon substitution $R\rightarrow S$. The operator $O_g$ is computed in the
Appendix by acting with $\Omega$ directly on fields $a$. Here we will use a different procedure which makes clear the connection to the calculation of $H_{RFT}$ in the previous section.
 Inserting resolution of identity in eq. (\ref{si}) we can write
\begin{equation}\label{qs}
O_g(k,\eta)\,=\,{1\over 2\pi}\sum_{n=1}^\infty \int_{x,y}\,e^{i\,k\,(x-y)}\,\int {1\over (2\pi)^{n-1}}\,
\Pi_{i=1}^{n-1}\,d^2z_i\,{d\eta_i}\ 
n\  Q_n^\dagger[S;z_i,\eta_i;x,\eta]\,Q_n[S;z_i,\eta_i;y,\eta]
\end{equation}
where $Q_n[S]$ is obtained from the amplitude $Q_n$ by substituting $S$ for $R$.
In the rest of this section we will simply use $Q_n$ to denote the amplitudes that enters eq.(\ref{qs}), but will keep in mind that they are related to the amplitudes used in the previous subsection by the aforementioned substitution. Note that now $Q_n$ are operators on the target Hilbert space and thus the averaging over the target as in eq.(\ref{gav}) is necessary.

The expression in eq.(\ref{qs}) can be calculated using eqs.(\ref{nodd}, \ref{neven}). In practical terms we have to cut one index linking one of the $\Xi$'s and $\Xi^\dagger$ (or $\Xi$ and $\bar b-b$ etc.) in eq.(\ref{nodd}) and eq.(\ref{neven}) and take the free transverse coordinate of $\Xi$ equal to $x$, while that of $\Xi^\dagger$ equal to $y$. Each term can be cut in $n$ possible positions, and all those cuts have to be summed over.
Thus for odd $n$ we obtain
\begin{eqnarray}\label{nodd1}
\sum_{n=0}^\infty\int_{dz_i,\eta_i}&(2n+1)& Q^\dagger_{2n+1}(z_i,\eta_i,x,\eta)\,Q_{2n+1}(z_i,\eta_i,y,\eta)\ =\nonumber \\
&=&
2\sum_{n=0}^\infty\Bigg[\sum_{m=0}^n\,(\bar b-b)\,N^\dagger_\perp\, K^{\dagger-1}\,
[\Xi\,\Xi^\dagger]^m_y\ _x[\Xi\,\Xi^\dagger]^{n-m}\,K^{-1}\,N_\perp\,(\bar b-b)\ + \nonumber\\
&+&
\sum_{m=0}^{n-1}(\bar b-b)\,N^\dagger_\perp\, K^{\dagger-1}\,[\Xi\,\Xi^\dagger]^m\,\Xi_x\ _y\Xi^\dagger\,[\Xi\,\Xi^\dagger]^{n-m-1}\,K^{-1}\,N_\perp\,(\bar b-b)\Bigg] \ =\nonumber\\
&=&2\,(\bar b-b)\,N^\dagger_\perp\, K^{\dagger-1}\,[ 1-\Xi\,\Xi^\dagger]^{-1}_{\ y}\ 
{_x}[1-\Xi\,\Xi^\dagger]^{-1}\,K^{-1}\,N_\perp\,(\bar b-b)\ + \nonumber \\
&+&2\,(\bar b-b)\,N^\dagger_\perp\, K^{\dagger-1}\,[ 1-\Xi\,\Xi^\dagger]^{-1}\,\Xi_x\ _y\Xi^\dagger\,
[ 1-\Xi\,\Xi^\dagger]^{-1}\,K^{-1}\,N_\perp\,(\bar b-b)\ =  \nonumber\\
&=&2\,(\bar b-b)\,N^\dagger_\perp K_y\ _xK^\dagger\, N_\perp\,(\bar b-b)\,+\,(\bar b-b)\,N^\dagger_\perp\, E_y \ _xE^\dagger\,
 N_\perp\,(\bar b-b)
\end{eqnarray}
where we have used a shorthand notation $_xAb\equiv \int_y A(x,y)b(y)$; $bA_y\equiv \int_xb(x)A(x,y)$.

For even $n$ we obtain
\begin{eqnarray}\label{neven1}
\sum_{n=1}^\infty\int_{z_i,\eta_i}&2n&\ Q^\dagger_{2\,n}(z_i,\eta_i,x,\eta)\,Q_{2n}(z_i,\eta_i,y,\eta)\ = \nonumber \\
&=&\sum_{n=1}^\infty\Bigg[\sum_{m=0}^n(\bar b-b)\,N^T_\perp \,K^{T-1}\,\Xi^\dagger[\Xi\,\Xi^\dagger]^m_y \ _x[\Xi\,\Xi^\dagger]^{n-m}\,K^{-1}\,N_\perp\,(\bar b-b)\ + \nonumber\\
&+&
\sum_{m=0}^{n-1}(\bar b-b)\,N^T_\perp \,K^{T-1}\,[\Xi^\dagger\,\Xi]^m_x\ _y\Xi^\dagger \,[\Xi\,\Xi^\dagger]^{n-m-1}\,K^{-1}\,N_\perp\,(\bar b-b)
\,+\,h.c.(x\rightarrow y)\Bigg]\ = \nonumber\\
&=&\Bigg[(\bar b-b)\,N^T_\perp\, K^{T-1}\,[ 1-\Xi^\dagger\,\Xi]^{-1}_{\ x} \ _y\Xi^\dagger\,
[1-\Xi\,\Xi^\dagger]^{-1}\,K^{-1}\,N_\perp\,(\bar b-b)\ + \nonumber \\
&+&(\bar b-b)\,N^T_\perp \,K^{T-1}\,[ 1-\Xi^\dagger\,\Xi]^{-1}_{\ x} \ _y\Xi^\dagger\,[ 1-\Xi\,\Xi^\dagger]^{-1}\,K^{-1}\,N_\perp\,(\bar b-b)\,+\,h.c.(x\rightarrow y)\Bigg]\ = \nonumber\\
&=&2(\bar b-b)\,N^T_\perp\, E^*_y\ _x\,K^\dagger\, N_\perp\,(\bar b-b)\,+\,
2\,(\bar b-b)\,N^\dagger_\perp\, E_y\ _x\,K^T \,N^*_\perp\,(\bar b-b)
\end{eqnarray}
Adding eqs.(\ref{nodd1}) and (\ref{neven1}) and using eq.(\ref{iden2}) it is now straightforward to show that
\begin{eqnarray}\label{singlein}
 O_g[k,\eta]&=&{1\over \pi}\int_{x,y} \,e^{i\,k\,(x-y)}\,(\bar b-b)\,
\bar N^\dagger_{\perp\,y} \ _x\bar N_\perp\,(\bar b-b)\nonumber\\ 
&=&{1\over \pi}\int_{x,y}\, e^{i\,k\,(x-y)}\, (b_L-b_R\,S^\dagger)\,(1-l-L_L)_y \ _x(1-l-L_L)\,( b_L-S\,b_R)\,.
\end{eqnarray}
Note that the gluon emission operator $O_g$ is independent of the  rapidity of the gluon $\eta$,
 as it should be for a boost invariant plateau.

As a corollary we note that integrating eq.(\ref{singlein}) over the transverse momentum we obtain the expression for total multiplicity
\begin{equation}
{dn\over d\eta}\,=\,{1\over \pi}\,\langle\ (b_L-b_R\,S^\dagger)\,[1-l-L_L+L_Ll+lL_L]\,( b_L-S\,b_R)\ \rangle_{j,S}\,.
\end{equation}

This generalizes the results of \cite{kovtuch} for the inclusive gluon spectrum in p-A scattering. We can check explicitly that in the KLWMIJ limit eq.(\ref{singlein}) indeed reduces to the known result of \cite{kovtuch,Braunincl, Baier, diff}.
In the limit of dilute projectile as before we take 
\begin{equation}
L_L\rightarrow l; \ \ \ \ \ \ \ \ \ b_i^a(x)\rightarrow{1\over 2\pi} \int d^2z {(x-z)_i\over (x-z)^2}j^a(z), \ \ \ \ \ \ 
 b_{L\,i}^a(x)\rightarrow{1\over 2\pi} \int d^2z {(x-z)_i\over (x-z)^2}S^{ab}j^b(z)\,.
\end{equation}
In this limit then the two factors $1-2l= \delta^{ij}-2{k_ik_j\over k^2}$ cancel against each other.
We then obtain
\begin{equation}
{dn(\eta,k)\over d\eta}\,=\,
{1\over 4\pi^3}\int_{x,y,z,\bar z} e^{ik(x-y)}\,\langle\  j^b(z)\left[S^{ab}(z)-S^{ab}(y)\right]{(y-z)_i\over (y-z)^2} {(x-\bar z)_i\over (x-\bar z)^2} \left[S^{ac}(\bar z)-S^{ac}(x)\right]j^c(\bar z)\ \rangle_{j,S}
\end{equation}
which is the known result \cite{kovtuch,Baier, diff}.

Recently single inclusive gluon spectrum in nucleus-nucleus collision was discussed in \cite{raju}. Although the setup of \cite{raju} is somewhat different from ours, it is  possible to establish close correspondence between the two approaches. The procedure of 
\cite{raju} is the following. The single gluon inclusive spectrum is defined as
\begin{equation}
{dn(\eta,k)\over d\eta}=
{1\over \pi}\int_{x,y} e^{ik(x-y)}\,\langle\,{\cal A}_i^a(y)\,{\cal A}_i^a(x)\,\rangle_{P,T}
\end{equation}
where ${\cal A}$ is the solution of classical Yang-Mills equations of motion with initial condition corresponding to colliding sheets of color charge density (nuclei)\cite{herlar}.  The solution should be taken at asymptotically large time after the collision. The averaging is then done over initial conditions with separate weight functions for the projectile and target color charge distributions.
This procedure is somewhat implicit since the classical field ${\cal A}$ has to be found by numerical solution of the classical equations. 

Comparing this to our result, we see that  eq.(\ref{singlein}) indeed "`feels"' like the classical field. The  expression 
\begin{equation}\label{class}
{\cal A}\,=\,\bar N_\perp\,(\bar b-b)\,=\,(1-l-L_L)\,\left[b_L-S\,b_R\right]\,.
\end{equation}
 plays the role of the classical field at asymptotically large times after the collision. With this identification our formula becomes very similar to that of \cite{raju} except for the fact that our expression is rather more explicit. To evaluate eq.(\ref{singlein}) there is no need to solve dynamical equations of motions, but rather only to solve the static classical equations
at early time, which express the classical field $b$ in terms of the color charge density $j$. 
In fact it may be possible to avoid this step altogether if one can write down directly a weight function for $b$ as advocated in the fourth paper in \cite{JIMWLK}. This is an interesting question and we intend to come back to it in the future.

We also mention, that an expression for ${d n(\eta, k)\over d\eta}$ was suggested in \cite{Kovnuc}. The averaging over the target and projectile weight functionals was performed in \cite{Kovnuc} using the McLerran-Venugopalan model \cite{mv}. Since we are still not at the point where we can calculate the average of the expression (\ref{singlein}), we cannot compare our result with the suggestion of \cite{Kovnuc}. 

Ref.\cite{raju} also discusses the evolution of the single gluon spectrum with energy. We have so far discussed only the explicit form of the observable itself. The evolution in our approach however is very straightforward to understand. We will discuss this later on in this section.

\subsection{Double and multi gluon inclusive amplitudes.}
The same strategy can be used to calculate the double inclusive cross section as well as higher gluon number correlations. In this paper we only consider multi gluon observables where all the counted  gluons 
have rapidities not too far away from each other, so that it is not necessary to consider rapidity evolution between them.
 The double gluon cross section for two gluons with rapidities $\eta$ and $\xi$ and transverse momenta $k$ and $p$ is defined as
\begin{eqnarray}\label{doug}
&&{dn(\eta,k;\xi,p)\over d\eta\, d\xi}\,=\,{1\over (2\pi)^2}\,\langle\,\langle 0\vert\,\Omega^\dagger\, \hat S^\dagger\, \Omega \,
a^{\dagger a}_i(\eta,k)\,a^{a}_i(\eta, k)\,a^{\dagger a}_i(\xi,p)\,a^{a}_i(\xi,p)\,\Omega^\dagger \,\hat S \,
\Omega\,\vert 0\rangle\rangle_{j,S}\\
&&\ \ \ \ \  =\,{1\over (2\pi)^2}\int_{(x,\bar x;y,\bar y)}e^{ik(x-\bar x)+ip(y-\bar y)}
\langle\,\langle 0\vert\,\Omega^\dagger \,\hat S^\dagger \,\Omega \,a^{\dagger a}_i(\eta,x)\,a^{a}_i(\eta, \bar x)\,
a^{\dagger a}_i(\xi,y)\,a^{a}_i(\xi,\bar y)\,
\Omega^\dagger \,\hat S\, \Omega\,\vert 0\rangle\,\rangle_{j,S}\nonumber\\
&&\ \ \ \ \ =\,{1\over (2\pi)^2}\,\sum_{n=2}^\infty \int_{x,\bar x,y\bar y}\,e^{i\,k\,(x-\bar x)\,+\,i\,p\,(y-\bar y)}\,\int 
{1\over (2\pi)^{n-2}}\,
\Pi_{i=1}^{n-2}\,d^2z_i\,{d\eta_i}\ n\,(n-1)\nonumber \\ 
&&\ \ \ \ \ \ \ \ \ \ \ \ \ \ \ \ \ \ \ \ \ \ \ \ \ \ \ \ \ \ \ \ \ \ \ \ \ \ \ \times\ 
\langle\, Q_n^\dagger(S;z_i,\eta_i;x,\eta,y,\xi)\ Q_n(S;z_i,\eta_i;\bar x,\eta,\bar y, \xi)\,\rangle_{j,S}\,.\nonumber
\end{eqnarray}
The calculation is straightforward albeit fairly long. The leading contribution to the double inclusive amplitude is obviously $O({1\over\alpha_s^2})$. The result for this contribution is
\begin{eqnarray}\label{leaddouble}
{dn(\eta,k;\xi,p)\over d\eta\, d\xi}&=&{1\over \pi^2}\int_{(x,\bar x;y,\bar y)}e^{ik(x-\bar x)+ip(y-\bar y)}
\langle(\left[(\bar b-b)\bar N^\dagger_{\perp\bar x} \ _x\bar N_\perp(\bar b-b)\right]\left[(\bar b-b)\bar N^\dagger_{\perp\bar y} \ _y\bar N_\perp(\bar b-b)\right]\rangle_{j,S}\nonumber\\
&=&{1\over \pi^2}\int_{(x,\bar x;y,\bar y)}e^{ik(x-\bar x)+ip(y-\bar y)}\langle
\left[b_L-b_RS^\dagger)(1-l-L_L)_{\bar x} \ _{ x}(1-l-L_L)( b_L-Sb_R)\right]\times\nonumber\\
&& \ \ \ \ \ \ \ \ \ \ \ \ \ \ \ \ \ \ \ \ \ \ \ \ \ \ \ \ \ \ \ \ \ \ \ \ \ \ 
\left[b_L-b_RS^\dagger)(1-l-L_L)_{\bar y} \ _{ y}(1-l-L_L)( b_L-Sb_R)\right]\rangle_{j,S}\,. \nonumber \\
\end{eqnarray}
This again is the "classical" contribution. It is "disconnected" in the sense that it does not depend on either rapidity $\eta$ or $\xi$ and as an operator on the valence and target Hilbert spaces, is equal to the square of the single inclusive amplitude $ O_g^2$.
This result immediately generalizes to multi-gluon inclusive amplitudes defined as
\beq
{dn(\eta_1,k_1;...,\eta_n,k_n)\over d\eta_1...d\eta_n}\,=\,\int Dj\,DS\,W^P[j]\,  \Gamma_n(k_1,...k_n;\eta_1,...\eta_n)\, W^T[S]
\eeq
with
\beq
 \Gamma_n(k_1,...k_n;\eta_1,...\eta_n)\,\equiv\ {1\over( 2\pi)^n}\, \langle 0\vert\,\Omega^\dagger\, \hat S^\dagger\, \Omega\ 
 a^{\dagger a}_i(\eta_1,k_1)\,a^{a}_i(\eta_1, k_1)\, ...a^{\dagger b}_j(\eta_n,k_n)\,a^{b}_j(\eta_n, k_n)
 \ \Omega^\dagger\, \hat S \,\Omega\,\vert 0\rangle\,.
\eeq
Similarly to the double gluon case, $\Gamma_n$ can be expressed in terms of $Q_n$:
\begin{eqnarray}
 \Gamma_n(k_1,...k_n;\eta_1,...\eta_n)&=&{1\over (2\pi)^n}\,\sum_{m=n}^\infty \int_{x_1,\bar x_1,...,x_n,\bar x_n}\,
e^{i\,k_1\,(x_1-\bar x_1)\,+...+\,i\,k_n\,(x_n-\bar x_n)}\,\int 
{1\over (2\pi)^{m-n}}\,{n!\over m!}\,\nonumber \\ 
&\times&
\Pi_{i=1}^{m-n}\,d^2z_i\,{d\eta_i}\ 
 Q_m^\dagger(S;z_i,\eta_i;x_1,\eta_1,...,x_n,\eta_n)\ 
Q_m(S;z_i,\eta_i;\bar x_1,\eta_1,...,\bar x_n, \eta_n)\,.\nonumber
\end{eqnarray}
 To leading order in the coupling constant
\begin{equation}\label{lo}
\Gamma_n(k_1,...,k_n;\eta_1,...\eta_n)\ \rightarrow\ 
\Gamma_n^{LO}(k_1,...,k_n)\ \equiv\ O_g(k_1)...O_g(k_n)
\end{equation}
or equivalently
\begin{equation}
{dn(k_1,...,k_n)\over d\eta_1...d\eta_n}\,=\,
{1\over \pi^n}\int_{(x_1,\bar x_1;... x_n,\bar x_n)}e^{ik_1(x_1-\bar x_1)+...+ik_n(x_n-\bar x_n)}\langle\,
[{\cal A}(x_1)\cdot {\cal A}(\bar x_1)]...[{\cal A}(x_n)\cdot {\cal A}(\bar x_n)]\,\rangle_{j,S}
\end{equation}
with the "classical field" ${\cal A}$ defined in eq.(\ref{class}). 
Thus all these observables have the same structure as discussed in \cite{raju} with the identification of the classical field at asymptoticaly late times given by eq.(\ref{class}).
All these leading order contributions do not depend on the rapidities of emitted gluons, since the gluons in leading order are emitted independently. This is the same as in the leading order BFKL calculations \cite{BFKL}. This of course does not mean that the multi-gluon amplitude is simply the product of single gluon ones
\begin{equation}
{dn(k_1,...,k_n)\over d\eta_1...d\eta_n}\ \ne\  {dn(k_1)\over d\eta_1}...{dn(k_n)\over d\eta_n}\,.
\end{equation}
It is the averaging over the projectile valence  and the target Hilbert spaces that breaks the factorization,  
even though the equality holds on the level of operators eq.(\ref{lo}).


Beyond the leading order things become more interesting. In particular the independence of rapidity does not hold anymore. We will only consider here the double inclusive gluon amplitude in detail. The calculation for multi-gluon inclusive is very similar and not more illuminating.
At order $O({1\over\alpha_s})$ the double inclusive amplitude has a contribution
with a nontrivial dependence on the rapidity difference $\eta-\xi$. This is the new feature of our calculation which is not present neither in KLWMIJ nor in JIMWLK limits. It also has not been considered in \cite{raju}, as it is formally the next to 
leading order contribution. 
Similar correlations, however, arise in the BFKL at NLO.

To calculate the correlated contribution we return to the definition eq.(\ref{doug}). The mechanics of the calculation is very similar to the single gluon inclusive amplitude. We have to cut two links between $\Xi$'s and $\Xi^\dagger$'s in eq.(\ref{nodd}) and eq.(\ref{neven}), and settle the cut links with the free transverse coordinates and rapidities of the counted gluons. This generates terms of $O(1/\alpha_s^2)$, $O(1/\alpha_s)$ and $O(1)$. The former terms add up to the result quoted in eq.(\ref{leaddouble}). 
Terms of $O(1)$ we neglect and concentrate on the terms  $O(1/\alpha_s)$. Some of the terms in the resulting expression do not depend on the rapidities of the gluons. We do not have control over this type of terms. The reason is the following. As we have discussed in \cite{rft} we have not been careful with the subleading terms in $H_{RFT}$ related to relative ordering between $b$ and $\Lambda$. The terms arising from a change of ordering would lead to a "virtual correction" to the classical field $b$ of order $\delta b\sim O(g)$. These terms would also be present in the expression for $Q_n$ and thus would give contribution to our present calculation at $O(1/\alpha_s)$. However the structure of these terms is clearly the same as that of the leading $O(1/\alpha_s^2)$ terms, and thus will not depend on rapidity. 

We conclude that at present we do not have control over rapidity independent terms in $O(1/\alpha_s)$
but can unambiguously calculate the rapidity dependent terms to this order. We note that there is nothing that fundamentally prevents us from treating the ordering more carefully and thus calculating all subleading terms in the nucleus-nucleus scattering case. We choose not to do so in this paper since it is a separate question and deserves a careful treatment in its own right.   

We thus concentrate on the terms depending on the rapidity difference $\eta-\xi$. In principle therefore we should be calculating ${dn(k,p)\over d\eta d\xi d(\eta-\xi)}$, but rather than taking an extra derivative we will simply subtract any term we get which does no depend on $\eta-\xi$.
A straightforward calculation along the lines described above gives:
\begin{eqnarray}\label{doublein}
&&{dn(\eta,k;\xi,p)\over d\eta d\xi}|_{correlated}\,=\,{1\over 2\pi^2}\int_{(x,\bar x;y,\bar y)}e^{ik(x-\bar x)+ip(y-\bar y)}\nonumber\\
&&\ \ \ \ \ \ \times\,\langle\,\Big[(\bar b-b)\bar N^\dagger_{\perp\bar x} (E^\dagger E)_{(x,\eta;\bar y,\xi)} \ _y\bar N_\perp(\bar b-b)+(\bar b-b)\bar N^\dagger_{\perp\bar y}(E^\dagger E)_{(y,\xi;\bar x,\eta)} \ _x\bar N_\perp(\bar b-b)\nonumber\\
&&\ \ \ \ \ \ \ +\,(\bar b-b)\bar N^\dagger_{\perp\bar x}(K^\dagger E)_{( x,\eta;y,\xi)} \ _{\bar y}\bar N^*_\perp(\bar b-b)+(\bar b-b)\bar N^T_{ \perp x}(E^\dagger K)_{( \bar x,\eta;\bar y,\xi)} \ _{y}\bar N_\perp(\bar b-b)\nonumber\\
&&\ \ \ \ \ \ \ +\,(\bar b-b)\bar N^\dagger_{\perp\bar x}(K^\dagger E)_{( y,\xi;x,\eta)} \ _{\bar y}\bar N^*_\perp(\bar b-b)+(\bar b-b)\bar N^T_{\perp x}(E^\dagger K)_{( \bar y,\xi; \bar x,\eta)} \ _{y}\bar N_\perp(\bar b-b)\Big]\,\rangle_{j,S}\,. \nonumber \\
\end{eqnarray}
We stress again, that all expressions in this equation and below depend on the single gluon scattering matrix $S$, rather than the matrix $R$.

To simplify this expression further we note that the integral of $\bar N$ over rapidity is real, and therefore
$\bar N^*(\bar b-b)=\bar N(\bar b-b)$. Further we note that the matrix $E$ is pure imaginary. Finally we have to remember that the operator in eq.(\ref{doublein}) has to be averaged over the projectile and target wave functions. We will assume that the averaging weights are rotationally invariant, and thus the result must be invariant under $k\rightarrow-k;\ p\rightarrow -p$, which is equivalent to $(\bar x,\bar y\ \rightarrow \ x,y)$. Under this assumption eq.(\ref{doublein}) becomes
\begin{eqnarray}\label{doublein1}
&&{dn(\eta,k;\xi,p)\over d\eta d\xi}|_{correlated}\,=\,{1\over 2\pi^2}\int_{(x,\bar x;y,\bar y)}\ e^{ik(x-\bar x)+ip(y-\bar y)}\\
&&\times\,\langle\,\Big[(\bar b-b)\bar N^\dagger_{\perp\bar x}[E^\dagger E]_{(x,\eta;\bar y,\xi)} \ _y\bar N_\perp(\bar b-b)+(\bar b-b)\bar N^\dagger_{\perp\bar y}(E^\dagger E)_{(y,\xi;\bar x,\eta)} \ _x\bar N_\perp(\bar b-b)\nonumber\\
&&+\,(\bar b-b)\bar N^\dagger_{\perp\bar x}[E^\dagger (K-K^*)]_{( x,\eta;y,\xi)} \ _{\bar y}\bar N_\perp(\bar b-b)
+(\bar b-b)\bar N^\dagger_{\perp\bar x}[E^\dagger (K-K^*)]_{( y,\xi;x,\eta)} \ _{\bar y}\bar N_\perp(\bar b-b)\Big]\,\rangle_{j,S}\,.\nonumber
\end{eqnarray}
We further note that the operators $E^\dagger E$ and $E^\dagger(K-K^*)$ are both Hermitian and real, and therefore symmetric. 
Thus we finally get
 \begin{eqnarray}\label{doublein2}
&&{dn(\eta,k;\xi,p)\over d\eta d\xi}|_{correlated}={1\over \pi^2}\int_{(x,\bar x;y,\bar y)}e^{ik(x-\bar x)+ip(y-\bar y)}\\
&&\times\,\langle\ \Big[(\bar b-b)\bar N^\dagger_{\perp\bar x}[E^\dagger E]_{(x,\eta;\bar y,\xi)} \ _y\bar N_\perp(\bar b-b)+(\bar b-b)\bar N^\dagger_{\perp\bar x}[E^\dagger (K-K^*)]_{( x,\eta;y,\xi)} \ _{\bar y}\bar N_\perp(\bar b-b)\Big]\ \rangle_{j,S}\,.\nonumber
\end{eqnarray}
To get a more explicit expression we first note the following two integrals
\begin{eqnarray}
&&\int_\xi{1\over 1+Ae^{\eta-\xi}} {1\over 1+Be^{\lambda-\xi}} ={1\over 2}{{B\over A}e^{\lambda-\eta}+1\over {B\over A}e^{\lambda-\eta}-1}\ln\left[{B\over A}e^{\lambda-\eta}\right]+C_1\\
&&\int_\xi{1\over 1+Ae^{\eta-\xi}}\left[ {1\over 1-Be^{\lambda-\xi}+i\epsilon}+ {1\over 1-Be^{\lambda-\xi}-i\epsilon}\right] ={{B\over A}e^{\lambda-\eta}-1\over {B\over A}e^{\lambda-\eta}+1}\ln\left[{B\over A}e^{\lambda-\eta}\right]+C_2\nonumber
\end{eqnarray}
The terms $C_1$ and $C_2$ do not depend on the rapidity difference $\eta-\lambda$. They are both formally  logarithmically divergent when the integration over $\xi$ is unrestricted. They do depend in principle on $\eta+\lambda$. The dependence on the sum of rapidities however, is simply part of the evolution of this observable in the rapidity difference between the projectile and the measured gluons. We will discuss this evolution in the next subsection, but are not interested in it for the current discussion.
We thus will drop the terms $C_1$ and $C_2$ in the following since they do not depend on the rapidity difference.
We next use these integrals to calculate the products of matrices $E$ and $K$ that enter eq.(\ref{doublein1}):
\begin{eqnarray}\label{ek}
[E^\dagger E]_{(x,\eta; y,\lambda)}&=&\langle x\vert{e^{\lambda-\eta}+1\over e^{\lambda-\eta}-1}(\lambda-\eta)\vert y\rangle \\
&-&{1\over 2}\langle x\vert (1-2l)S(1-2l)(1-2L_R)
{{D^2_R\over \bar D^2}e^{\lambda-\eta}+1\over {D^2_R\over \bar D^2}e^{\lambda-\eta}-1}\ln\left[{D^2_R\over \bar D^2}e^{\lambda-\eta}\right]S^\dagger(1-2L_L)\vert y\rangle\nonumber\\
&-&{1\over 2}\langle x\vert (1-2L_L)S{{ D_R^2\over \bar D^2}e^{\eta-\lambda}+1\over { D_R^2\over \bar D^2}e^{\eta-\lambda}-1}\ln\left[{D_R^2\over \bar D^2}e^{\eta-\lambda}\right](1-2L_R)(1-2l)S^\dagger(1-2l)\vert y\rangle\nonumber\\
\left[E^\dagger (K-K^*)\right]_{(x,\eta; y,\lambda)}&=&2\langle x\vert{e^{\lambda-\eta}-1\over e^{\lambda-\eta}+1}(\lambda-\eta)\vert y\rangle \\
&-&\langle x\vert (1-2l)S(1-2l)(1-2L_R)
{ {D^2_R\over \bar D^2}e^{\lambda-\eta}-1\over {D^2_R\over \bar D^2}e^{\lambda-\eta}+1}\ln\left[{D^2_R\over \bar D^2}e^{\lambda-\eta}\right]S^\dagger(1-2L_L)\vert y\rangle\nonumber\\
&-&\langle x\vert (1-2L_L)S
{ {D^2_R\over \bar D^2}e^{\eta-\lambda}-1\over {D^2_R\over \bar D^2}e^{\eta-\lambda}+1}\ln\left[{D^2_R\over \bar D^2}e^{\eta-\lambda}\right]
(1-2L_R)(1-2l)S^\dagger(1-2l)\vert y\rangle\,.\nonumber
\end{eqnarray}
There clearly is a nontrivial dependence on the rapidity difference in these expressions. Interestingly enough this dependence does not disappear even when the rapidities are far from each other.
At large values of rapidity difference $\lambda-\eta$ we have
\begin{eqnarray}\label{large}
&&2[E^\dagger E]_{(x,\eta; y,\lambda)}|_{|\lambda-\eta|\gg 1}=\left[E^\dagger (K-K^*)\right]_{(x,\eta; y,\lambda)}|_{|\lambda-\eta|\gg 1}\ = \\
&&\langle x\vert 2-(1-2l)S(1-2l)(1-2L_R)S^\dagger(1-2L_L)-(1-2L_L)S(1-2L_R)(1-2l)S^\dagger(1-2l)\vert y\rangle |\lambda-\eta|\,.
\nonumber
\end{eqnarray}
Thus we find that the rapidity correlation between the two gluons in the next to leading order does not disappear at large rapidities. Obviously we cannot use these expressions when the rapidity difference is too large, since then we have to take into account the rapidity evolution between $\eta$ and $\lambda$. Still $\eta-\lambda$ can be taken parametrically of order one, but numerically greater than one. In this regime our calculation should be valid and the effect is visible. 

It would be very interesting to understand the physics of the appearance of these correlations and their possible implications. It is rather clear that we should understand some of these terms as the first correction due to the rapidity evolution between $\eta$ and $\lambda$, which indeed should be a formally subleading correction to the leading result eq.(\ref{doublein}). At the moment we cannot make any further comments on the subject and leave it as a question well worth studying. 

We note however, that not all the terms in eq.(\ref{ek}) can be attributed to the evolution. Some of them, when integrated over rapidity, contribute directly to the leading order RFT Hamiltonian. The Hamiltonian can be represented in terms of the inclusive gluon amplitudes $\Gamma_n$
in the manner analogous to the representation in terms of amplitudes $Q_n$
\begin{equation}
H_{RFT}\,=\,\Sigma_{n=1}^\infty {(-1)^n\over n!}\,{d\over d Y}\,\int_{\{k_i; \eta_i\}}\ 
\Gamma_n(k_1,...,k_n,\eta_1,...\eta_n)|_{Y=0}\,.
\end{equation}
This picks out the linear in $Y$ piece in the integral of $\Gamma_n$, including of course $\Gamma_2$, over rapidities. Such a contribution does not come from the leading order piece, which is clearly proportional to $Y^2$ (for $\Gamma_2$) nor from the $\eta-\xi\gg 1$ integration region of eq.(\ref{doublein1}), since it is proportional to $Y^3$ (see eq.(\ref{large})). However integrating eq.(\ref{doublein1}) over the two rapidities one will also pick a contribution proportional to first power of $Y$ from the region $\eta-\xi\sim 1$ where the two gluons are locally correlated in rapidity over and above the long range linear rapidity correlation explicit in eq.(\ref{large}). Thus for small rapidity differences eq.(\ref{doublein1}) contains local physical effects not related to the rapidity evolution between $\eta$ and $\xi$.

Quite generally existence of such short range correlations can be inferred from the structure of our calculation of $H_{RFT}$. Recall that $H_{RFT}$ is obtained by taking the (sum of the) squares of exclusive $n$-gluon production amplitudes $Q_n$, integrating them over the rapidities of all the gluons, and picking the piece of the integral that is linear in the length of the total rapidity interval $Y$. Such a linear in  $Y$ piece naturally arises from the integral of $Q_1^\dagger Q_1$, since only one rapidity variable is integrated over. However for $Q^\dagger_nQ_n$ with $n>1$ the only way such a linear piece can arise is if there is finite excess (or depletion) of probability for {\it all} $n$ gluons to be correlated in rapidity within a finite interval $\eta_1\sim\eta_2\sim...\sim\eta_n$ over a totally uncorrelated situation. Since as we have seen earlier all $Q^\dagger_nQ_n$ with $n\ge 1$ contribute to $H_{RFT}$, it means that such short range correlations are indeed present. Eq.(\ref{doublein1}) is just a specific example of this correlation.

We have not attempted to study these local correlations in any detail. 
It is tempting to speculate, however, that they lead to anti-bunching, rather than bunching, thereby depleting the probability to emit several gluons close to each other in rapidity. This would
then impose sort of a "rapidity veto". Such an effect appears in the next to leading order perturbative approaches \cite{pertveto} and has been used in the framework of the nonlinear high energy evolution to emulate energy conservation \cite{evolveto}. It would be interesting if the inclusion of Pomeron loop effects discussed here and in \cite{rft} implements such a veto automatically.

We also note that extending our results to include evolution between the rapidities of the observed gluons, along the lines of \cite{multig} is fairly straightforward and we plan to address this question in near future. It would be very interesting to compare this to the approach of the last paper in \cite{raju}.

\subsection{Evolution with rapidity}
As we have already stated, we are not going to discuss the evolution with respect to rapidity differences between the counted gluons. On the other hand the evolution with respect to the total energy is covered by our present derivation in a trivial manner. All the multi-gluon observables discussed so far have the form (\ref{gav})
\begin{equation}\label{obs}
\bar O(y,Y)\,=\,\langle \, O(j,S)\,\rangle_{j,S}\,\equiv\,\int[dj][dS] \,W^P_{y}[j]\  O(j,S)\ W^T_{Y-y}[S]
\end{equation}
where $O(j,S)=\Gamma_n$ and $W^P_y$ is the weight functional for averaging over the projectile degrees of freedom, while $W^T_{Y-y}$ is
the same for target degrees of freedom. The total rapidity of the process is $Y$ and the gluons are measured at rapidity $y$ away from the projectile.
This expression automatically has a factorized structure discussed in \cite{raju}, independently whether the scattering objects are nuclei or dipoles. 
One can of course ask how does this observable evolve with any of the two rapidity variables it depends on: $y$ or $Y-y$; or in fact with $Y$ at fixed $y$ or $Y-y$.
The answer to this question is straightforward given that we know the evolution of the weight function $W^P$ 
derived in the previous section. Thus for example the evolution with respect $y$ at fixed $Y-y$ is given by
\begin{equation}\label{evp}
{\partial\over \partial y}\,\bar O(y,Y-y)|_{Y-y}\,=\,-\,\int[dj][dS] \Big[H_{RFT}[U,R]\,W^P_{y}[j]\Big]\  O(j,S)\ W^T_{Y-y}[S]
\end{equation}
with $H_{RFT}$ given in eq.(\ref{hg}).
We have not derived directly the evolution of the target weight functional. However as was shown in \cite{klwmij},\cite{something}, Lorenz invariance requires the target weight function to evolve with the dual Hamiltonian $H_{RFT}[R_S,S]$ where $R_S$ is to $S$, what $R$ is to $U$.  Namely 
\beq
j^{T\,a}\,\equiv\,-{1\over g}\,f^{abc}\,\partial_i\,S^{\dagger bd}\,\partial_i \,S^{dc}\,; \ \ \ \ \ \ \ \ \ \ \ \ \ \ \ \ \ \ 
R_S(x)\ =\ {\cal P}\,\exp\{g\int_0^1 dt\  T^c\ {\delta\over\delta j^{T\,c}(x,t)}\}\,.
\eeq
where $j^T$ is the color charge density of the target.
Using this we can write the evolution with respect to $Y$ at fixed $y$ as
\begin{equation}\label{evt}
{\partial\over \partial Y}\,\bar O(y,Y-y)|_{y}\,=\,-\,\int[dj][dS] \ W^P_{y}[j]\  O(j,S)\ \Big[H_{RFT}[R_S,S]\,W^T_{Y-y}[S]\Big]
\end{equation}
It was also shown in \cite{something} that the complete $H_{RFT}$ must be self dual, namely $H_{RFT}[U,R]=H_{RFT}[R,U]$. We have not verified explicitly that $H_{RFT}$ of eq.(\ref{hg}) satisfies the property of self duality. As noted in \cite{rft}, the technical issue that has to be resolved before we can address this question is the duality transformation properties of $U_L$. This is an  interesting question which we plan to address in future.  Assuming self duality of $H_{RFT}$,  eq.(\ref{evt}) can be also written as
\begin{equation}\label{evt1}
{\partial\over \partial Y}\,\bar O(y,Y-y)|_{y}\,=\,-\,\int[dj][dS]\  W^P_{y}[j]\ O(j,S)\ \Big[H_{RFT}[S,R_S]\,W^T_{Y-y}[S]\Big]
\end{equation}
Combining eqs.(\ref{evp}) and (\ref{evt}) one can also write evolution equation with respect to
other combinations of $y$ and $Y$. Thus if we want to follow the evolution of $O$ with rapidity $y$ keeping the total rapidity fixed we have
\begin{eqnarray}\label{evo}
{\partial\over \partial y}\,\bar O(y,Y-y)|_Y&=&-\,\int[dj][dS]\ \Bigg\{\Big[H_{RFT}[U,R]\,W^P_{y}[j]\Big]\  O(j,S)\ 
W^T_{Y-y}[S]\nonumber \\
& -&\  W^P_{y}[j]\ O(j,S)\ \Big[H_{RFT}[S,R_S]\,W^T_{Y-y}[S]\Big]\Bigg\}
\end{eqnarray}
The factorized structure of this evolution is the same as discussed earlier for dipole-nucleus scattering in \cite{kovtuch},\cite{multig} and nucleus-nucleus scattering in \cite{raju}. It is rather universal, and one might even say trivial. It does not depend on the nature of colliding objects and, by definition,  is the generic property of any observable that can be represented in the form eq.(\ref{obs}). Of course, not any observable has this representation. Examples to the contrary are diffractive observables or observables nonlocal in rapidity \cite{diff},\cite{multig}.

The details of the evolution on the other hand depend on the situation.
In the dipole-nucleus scattering \cite{kovtuch},\cite{multig} one does not have to keep the full $H_{RFT}$ in the evolution equation. Instead the Hamiltonian which acts on the projectile weight function can be taken as $H_{KLWMIJ}[j,R]$. In the nucleus-nucleus scattering, according to \cite{raju} the full Hamiltonian reduces to $H_{JIMWLK}[U,\delta/\delta j]$. The natural question in this context is, does our formula eq.(\ref{evo}) allow to truncate $H_{RFT}$ to the JIMWLK form when considering multi-gluon production in nucleus-nucleus scattering. We will discuss this question along with some other points in the next section.

\section{Discussion.}
The main result of the present paper is the explicit expression for the single gluon and multi-gluon production amplitudes which include Pomeron loop effects. To summarize,
in the leading order in $\alpha_s$, the n-gluon production amplitude is given by a "semiclassical" expression
\begin{equation}
{dn(k_1,...,k_n)\over d\eta_1...d\eta_n}={1\over \pi^n}\int_{(x_1,\bar x_1;... x_n,\bar x_n)}e^{ik_1(x_1-\bar x_1)+...+ik_n(x_n-\bar x_n)}\langle
[{\cal A}(x_1)\cdot {\cal A}(\bar x_1)]...[{\cal A}(x_n)\cdot {\cal A}(\bar x_n)]\rangle_{j,S}
\end{equation}
where the "classical field" ${\cal A}$ is an explicit function of the projectile and target fields
\begin{equation}\label{f3}
{\cal A}\ =\ \bar N\,(\bar b-b)\ =\ (1-l-L_L)\ \left[b_L-S\,b_R\right]\,.
\end{equation}
This result reproduces the known expressions for the "dipole-nucleus" scattering. It has the same general structure as the expression for nucleus-nucleus scattering of \cite{raju}. The important difference between our result and the expressions in \cite{raju} is that our expressions for the observables are explicit functions of the projectile and target fields, while the procedure of \cite{raju} involves further numerical solution of dynamical equations of motion to determine the Yang-Mills fields at asymptotically late times. In this respect our expression eq.(\ref{f3}) is the explicit solution of the classical equations of motion, needed to apply the formalism of \cite{raju}. It would indeed be very interesting to check explicitly by numerical simulation whether the numerical solution of Yang-Mills equations after the collision does reproduce eq.(\ref{f3}). We believe this should be the case, once the difference between the gauge fixing conditions used in \cite{raju}
  and in the present paper is accounted for. We note in this connection that the eikonal approximation in the symmetric gauge/frame does indeed lead to solution of classical equations as the leading perturbative contribution\cite{eikon}. Thus we believe that our use of the eikonal approximation does not make the accuracy of our approach different from that of \cite{raju}.

We have also shown that two - and higher gluon amplitudes have subleading correlated pieces where all gluons at adjacent rapidities are emitted in a correlated way.  As we have discussed above, the logarithmic part of this correlation is likely to be the first correction due to the rapidity evolution between the emitted gluons, while   
the short range correlation, with the correlation length of order unity, is not an evolution effect. These are gluons that upon integration over rapidity contribute directly to $H_{RFT}$. We suspect that these correlations are in fact negative, and thus manifest themselves as a sort of rapidity veto due to the Pomeron loop effects.

Finally we have shown that the multi-gluon observables evolve with rapidity according to $H_{RFT}$ derived in \cite{rft}.
In this context we again have to discuss the relation of our results to those of \cite{raju}.
Ref.\cite{raju} discusses the evolution of the single and multi gluon production amplitudes with rapidity. The rapidity that changes could be either the rapidity of the measured gluons at fixed total energy (evolution with $y$ at fixed $Y$), or the total rapidity of the process keeping the gluons at mid-rapidity (evolution with $Y$ and $y$ so that $Y=2 y$). In all cases ref. \cite{raju} asserts that the observables should be evolved with the JIMWLK Hamiltonian. Can we reconcile the two statements? Consider for example the evolution
\begin{equation}
{\partial\over \partial y}\,{\partial n\over \partial \eta}|_{Y-y}\ =\ -\,
\int[dj][dS] \,\Big[H_{RFT}[U,R]\,W^P_{y}[j]\Big]\ O_g(j,S)\ W^T_{Y-y}[S]
\end{equation}
with the observable $O_g$ being the single gluon amplitude $O_g={\cal A}{\cal A}$ (\ref{si}). 
Since $H_{RFT}$ is a hermitian operator, we can also write
\begin{equation}
{\partial\over \partial y}\,{\partial n\over \partial \eta}|_{Y-y}\ =\ 
-\,\int[dj][dS]\, W^P_{y}[j]\, \Big[H_{RFT}[U,R]\ O_g(j,S)\Big]\ W^T_{Y-y}[S]
\end{equation} 
Let us now consider the question whether we can expand the operator $R(x)=\exp\{gT^a{\delta\over\delta j^a(x)}\}$ in powers of $\delta/\delta j$ when acting on the observable $O_g$. As is clear from eq.(\ref{f3}), the observable $O_g$ is a function of $gb$ and $g\bar b$ only. The fields $gb$ and $g\bar b$ in turn are functions of $gj$, as is obvious from the classical equations through which they are defined. Thus we have
\begin{equation}
{\delta\over \delta j}\,O_g\,\sim\, g{\delta (gb)\over \delta (gj)}\,{\delta\over \delta (gb)}\,O_g
\end{equation}
Thus each term in the expansion of $R$ when acting on the operator $O_g$ brings a factor of $\alpha_s$. The same argument goes through for any leading order multi-gluon inclusive amplitude. On the basis of this argument we therefore can conclude that for this particular set of observables, one can expand the Hamiltonian $H_{RFT}$ to leading order in $\delta/\delta j$. As we know, the leading order in this expansion is $H_{JIMWLK}$. Thus, on the face of it, we recover the conclusion of \cite{raju} - the evolution hamiltonian for these observables can be taken as $H_{JIMWLK}$. 

The previous line of argument immediately begs two questions.  First, how confident are we in the robustness of the argument itself. 
The answer to this question is not completely clear to us. Although naively the argument looks correct, there are examples of  situations where a similar argument fails. Consider the evolution of the dipole-dipole scattering amplitude using the KLWMIJ kernel.
Ignoring the details irrelevant to the present discussion we can write the evolution equation for the $S$ - matrix as
\beq\label{dev}
{d\over dY}{\cal S}\ =\ -\,\int d j d\alpha \, W^P[j]\ \  {\cal K}(x,y,z)\,j_x\,(1\,-\,R)_zj_y\,\ \ e^{i\int_x\,j(x)\,\alpha(x)}\,W^T[\alpha]\,.
\eeq
where $\cal K$ is the kernel appearing in $H_{KLWMIJ}$ (\cite{klwmij}).
Since the target is a dipole, the field $\alpha\sim O(g)$. It then follows by the argument similar to the one given above that when acting on this observable ($e^{i\int_x\,j(x)\,\alpha(x)}$), 
the $R$'s in the KLWMIJ Hamiltonian can be expanded to leading order in  $\delta/\delta j$. The leading order expansion is simply the BFKL Hamiltonian. Thus we would conclude that if the target is a dipole, we can always use the BFKL Hamiltonian rather than the full KLWMIJ. However we know that this is not the case. The BFKL Hamiltonian can be used only for evolution to rapidity of order $Y\sim {1\over \alpha_s}\ln {1\over \alpha^2_s}$.
Further evolution with BFKL Hamiltonian violates unitarity of the amplitude, while the full dipole or KLWMIJ Hamiltonian preserves unitarity. Thus even though the formal argument about expansion of $R$ at every step of evolution can be made, the cumulative effect of evolution to large enough rapidity is such that the expanded Hamiltonian misses a very important physical effect which leads to qualitative change in the evolution. Note that if the projectile is not a dipole, but rather a "nucleus" - a state with large dipole density in its wave function, the breakdown of $H_{BFKL}$ happens much earlier, at $Y\sim {1\over \alpha_s}$\footnote{We do not claim that the nuclear weight function can be evolved with $H_{KLWMIJ}$. Physically of course this does not make much sense, since the nonlinearities in the wave function are important. We merely point out that if one does it as a mathematical exercise, the formal argument about expanding $R$'s breaks down rather quickly.}. In this case we can point out to a distinct physical effect that is missed by the expansion - multiple scattering corrections due to scattering of more than one dipole of the projectile on the target. Those corrections are important if the projectile contains many dipoles, even if the target is dilute. For a dipole projectile multiple scattering corrections become important later in the evolution, when the dipole wave function becomes dense, while for a nucleus this happens much earlier.  In the case of the evolution of multi-gluon amplitudes in the nucleus-nucleus scattering discussed above, we do not have similar understanding. Still we think that one has to take the argument with a grain of salt. It makes perfect sense to ask how far one can evolve these observables in rapidity without encountering a problem of missing some important physical effect. We thus would like to advocate caution on this issue. The problem in our view needs to be studied further. 

The second question is this. If the argument is indeed correct, does this mean that we can ignore the difference between $H_{JIMWLK}$ and $H_{RFT}$ for all observables, and simply not bother with any of the calculations in \cite{rft}? 
The answer to this is clearly negative. The fact that the evolution simplifies for a certain set of observables, does not mean that it simplifies for all interesting observables. A simple example of a similar situation is the dipole-nucleus scattering. In this case the weight function of the projectile dipole $W^P[j]$ evolves according to $H_{KLWMIJ}$. This evolution is significantly different from $H_{BFKL}$, since as is well known $H_{KLWMIJ}$ leads to unitarization of the scattering amplitude, while $H_{BFKL}$ does not. On the other hand, if we consider a single inclusive gluon amplitude, the observable itself is quadratic in $j$ (see \cite{kovtuch}, \cite{diff}). Thus when acting on this observable, $H_{KLWMIJ}$ and $H_{BFKL}$ are identical. The fact that one cannot use $H_{BFKL}$ to evolve the weight functional $W^P[j]$ is another way of saying that there are some interesting observables on which the action of $H_{KLWMIJ}$ and $H_{BFKL}$ is not equivalent. As mentioned above, one of such observables is the forward scattering amplitude, which is unitarized by the KLWMIJ evolution but not by BFKL.
Other examples of such observables include various diffractive amplitudes \cite{diff} and also multi-gluon inclusive amplitudes in the case when the rapidity differences between observed gluons are large and the evolution between them has to be taken into account \cite{multig}.

We expect the situation to be similar for the case at hand. Clearly, $R$ in $H_{RFT}$ cannot be expanded when we calculate the forward scattering amplitude on a nuclear target. In this case the observable is $\exp\{ij\alpha\}$. When acting on it, $R$ becomes the matrix $S$, which is not perturbatively close to unity. Thus expansion is not possible.

We expect that multi-gluon amplitudes with large rapidity differences also do not allow expansion of $R$. The observables associated with these amplitudes are calculable and we hope to present results of this calculation soon. However, we can find a hint that all derivatives in $R$ are important, by examining the results of the previous section.
Consider the correlated term in the double gluon inclusive amplitude. We expect that the evolution between the rapidities of the two gluons, when the rapidity difference is large, should be given by $H_{RFT}$, or at least closely linked to it. On the other hand, as we have discussed above, the first term in this evolution is likely just the "long range" part of the rapidity correlated term eq.(\ref{doublein2}), that is eq.(\ref{large}). 
Examining eq.(\ref{large}) we see that it is just the first order expansion of the denominator of $H_{RFT}$ eq.(\ref{hg}) around $(1-2l)R^\dagger(1-2l)$. These terms would not depend on $S$ if we were to truncate $H_{RFT}$ at $H_{JIMLK}$, but they clearly depend on $S$ in eq.(\ref{large}). This suggests that the factors $R$ cannot be expanded and the complete Hamiltonian $H_{RFT}$ is important in the evolution of this observable.
We note that the conclusion of ref.\cite{raju} is different, namely that $H_{JIMWLK}$ is adequate also for evolution of multi-gluon observables with large rapidity differences. We feel therefore that the question warrants further study.

Finally we want to mention that it would be very interesting to understand how to perform the averaging over the valence and target fields. This would allow one to calculate physical observables like multi-gluon spectra. One can in principle use the McLerran-Venugopalan model \cite{mv} to specify the projectile and target weight functionals, at least to get a qualitative idea about the behavior of the observables. However even then one has to understand how to calculate correlators of three (dependent) matrix degrees of freedom $U_R,\ U_L,\ S$, which is far from trivial. This can certainly be done numerically, but one would like to be able to understand at least the basics analytically. At the moment this is an open question.

 \section*{Acknowledgments}

We are grateful to Javier Peressutti for his participation in the initial stages of this work.  We thank Francois Gelis and Raju Venugopalan  for several interesting discussions.
TA and AK acknowledge support from the DOE through the grant DE-FG02-92ER40716. The work 
of ML is partially supported by the DOE grants DE-FG02-88ER40388 and DE-FG03-97ER4014.

\appendix

\section{Appendix -  Single gluon inclusive amplitude.}\label{sec:A}
In this appendix we rederive the expression for the single gluon inclusive amplitude without the use of the amplitudes $Q_n$,
 but instead working directly from the definition.
\begin{equation}
\hat O_g\,=\,
{1\over 2\pi}\,\langle 0\vert\,\Omega^\dagger\, \hat S^\dagger\, \Omega \,
a^{\dagger a}_i(\eta,k)\,a^{a}_i(\eta, k)\,\Omega^\dagger\, \hat S\, \Omega\,\vert 0\rangle\,.
\end{equation}
In the parametric range we are interested in, namely when both the colliding objects carry fields of order $1/g$, we expect the number of produces gluons to be of order $1/\alpha_s$. We will calculate the single gluon spectrum only to this order. Let us first of all calculate
\begin{equation}
\Omega^\dagger \,\hat S^\dagger\, \Omega\, a_\alpha\,\Omega^\dagger \,\hat S^\dagger \,\Omega
\end{equation}
where now the index $\alpha$ stands for all discrete indices as well as momenta. In this equation and the following we assume that the rapidity of the gluon operator $a(\eta,k)$ is in the infinitesimal rapidity bin created by the operator $\Omega$. Thus in practice we are calculating directly the derivative with respect to rapidity. 
Recall that
\begin{equation}
\Omega\,=\,{\cal C}\,{\cal B}\,.
\end{equation}
Thus
\begin{equation}
\Omega a_i\Omega^\dagger={\cal C}\,{\cal B}\,
a_\alpha {\cal B}^\dagger\,{\cal C}^\dagger\,=\,{\cal C}\,[\Theta_{\alpha\beta}a_\beta+\Phi_{\alpha\beta}a^\dagger_\beta]\,
{\cal C}^\dagger\,.
\end{equation}
When acting by the operator $\cal C$ we keep in mind that we are only interested in the leading order contribution. Thus we do not need to worry about the action of the operator $\cal C$ on $j$ in $\Theta$ and $\Phi$. Also recall that 
\begin{equation}
{\cal C}\,a_\alpha \,{\cal C}^\dagger\,\approx\, a_\alpha\,-\,\sqrt 2\,i\,b_\alpha 
\end{equation}
In this relation we have neglected terms which are themselves of order one, and are proportional to $A(x^-=0)$. The reason is that we only need to keep terms, which by subsequent application of another operator ${\cal C}$ can be shifted by $b$, thus generating terms of order $1/g$. However $A(x^-=0)$ commutes with ${\cal C}$ and thus will not generate such contributions. Thus to the required order
\begin{equation}
\Omega \,a_\alpha\,\Omega^\dagger\,=\,
\Theta_{\alpha\beta}\,a_\beta\,+\,\Phi_{\alpha\beta}\,a^\dagger_\beta\,-\sqrt 2\,i\, (\Theta-\Phi)_{\alpha\beta}\,b_\beta[J]\,.
\end{equation} 
The subsequent action of the operator $\hat S$ simply rotates all creation/annihilation operators as well as all the charge density operators by the single gluon scattering matrix $S$
\begin{equation}
\hat S^\dagger\,\Omega\, a_\alpha\,\Omega^\dagger \,\hat S\,=\,
\Theta_{\alpha\beta}[SJ]\,S_{\beta\gamma}\,a_\gamma\,+\,\Phi_{\alpha\beta}[SJ]\,S_{\beta\gamma}\,
a^\dagger_\gamma\,-\,\sqrt 2\,i\,N_{\alpha\beta}[SJ]\,b_\beta[SJ]\,.
\end{equation}
Finally we apply again the transformation with the operator $\Omega$. The only relevant part of this transformation is the action of the operator $\cal C$. All the rest, as before does not give a leading order contribution
\begin{eqnarray}
\Omega^\dagger \hat S^\dagger \Omega a_\alpha\Omega^\dagger \hat S^\dagger \Omega&=&\sqrt 2i\Theta_{\alpha\beta}[SJ]S_{\beta\gamma}b_\gamma-\sqrt 2i\Phi_{\alpha\beta}[SJ]S_{\beta\gamma}b_\gamma-\sqrt 2iN_{\alpha\beta}[SJ]b_\beta[SJ]\nonumber\\
&=& \sqrt 2\,i\,N_{\alpha\beta}[SJ]\,\Big\{S_{\beta\gamma}\,b_\gamma[J]\,-\,b_\beta[SJ]\Big\}\,.
\end{eqnarray}
A similar relation holds for $a^\dagger$. Collecting this together we obtain for the single gluon inclusive spectrum
\begin{equation}
\hat O_g\,=\,{1\over\pi}\,\int_{x,y,z,\bar z}\,e^{ik(z-\bar z)}\,\bigg\{Sb[J;x]-b[SJ;x]\bigg\}\,
N^\dagger[SJ; x,z]\,N[SJ; \bar z,y]\,\bigg\{Sb[J;y]-b[SJ;y]\bigg\}
\end{equation}
which is the same as derived in the body of the paper (eq. \ref{singlein}).

%

\begin{thebibliography}{99}

\bibitem{rft} T. Altinoluk, A. Kovner, M. Lublinsky and J. Peressutti, arXiv:0901.2559.


\bibitem{gribov}   V.~N.~Gribov,
 Sov.\ Phys.\ JETP {\bf 26}, 414 (1968)
 [Zh.\ Eksp.\ Teor.\ Fiz.\  {\bf 53}, 654 (1967)];\\
  J.~Bartels,
  Nucl.\ Phys.\ B {\bf 151}, 293 (1979).
 J.~Bartels, Z.Phys. C60, 471, 1993; J.~Bartels and M.~Wusthoff, Z. Phys. {\bf C 66}, 157, 1995; J.~Bartels, Z.Phys. {\bf C60}, 471 (1993);
J.~Bartels and M.~Wusthoff, Z. Phys. {\bf C66}, 157 (1995);
 J.~Bartels and C.~Ewerz, JHEP  {\bf 9909}, 026 (1999).
e-Print: hep-ph/9908454;  J.~Bartels, L.~N.~Lipatov and M.~Wusthoff,
  Nucl.\ Phys.\ B {\bf 464}, 298 (1996).
  e-Print: hep-ph/9509303;
  M.~A.~Braun and G.~P.~Vacca,
  Eur.\ Phys.\ J.\ C {\bf 6}, 147 (1999).
 e-Print: hep-ph/9711486;
M. Braun  Phys. Lett. {\bf B483}, 115 (2000).
e-Print: hep-ph/0003004;

  
  
  
  


\bibitem{balitsky} I. Balitsky, { Nucl. Phys.}  {\bf B463}, 99 (1996); 
e-Print: hep-ph/9509348;
{ Phys. Rev. Lett.} {\bf 81} 2024 (1998);  e-Print: hep-ph/9807434;
{Phys. Rev.}{\bf D60} 014020 (1999); e-Print: hep-ph/9812311.

\bibitem{JIMWLK} J. Jalilian Marian, A. Kovner, A.Leonidov and H.
Weigert,
{ Nucl. Phys.}{\bf  B504} 415 (1997);  e-Print: hep-ph/9701284
{ Phys. Rev.} {\bf D59} 014014 (1999);  e-Print: hep-ph/9706377
J. Jalilian Marian, A. Kovner and H. Weigert, {Phys. Rev.}{\bf D59} 
014015 (1999); 
e-Print: hep-ph/9709432;
A. Kovner and J.G. Milhano, { Phys. Rev.} {\bf D61} 014012 (2000); e-Print Archive: hep-ph/9904420.
 A. Kovner, J.G. Milhano and H. Weigert,
{Phys. Rev.} {\bf D62} 114005 (2000); e-Print: hep-ph/0004014;
 H. Weigert, { Nucl. Phys.} {\bf A 703} (2002) 823; e-Print: hep-ph/0004044;
 




\bibitem{Kovchegov}
  Y.~V.~Kovchegov,
  Phys.\ Rev.\ D {\bf 61}, 074018 (2000).
e-Print Archive:hep-ph/9905214].
  
\bibitem{cgc}  E.Iancu, A. Leonidov and L. McLerran, {\it Nucl. Phys.} 
{\bf A 692}, 583 (2001); e-Print: hep-ph/0011241; {Phys. Lett.} {\bf B
510}, 133 (2001); e-Print: hep-ph/0102009; 
E. Ferreiro, E. Iancu, A. Leonidov, L. McLerran;  
{Nucl. Phys.}{\bf A703}, 489 (2002); e-Print: hep-ph/0109115;



\bibitem{klwmij}  A. Kovner and M. Lublinsky, Phys.\ Rev.\ D {\bf 71}, 085004 (2005), e-Print: hep-ph/0501198.


\bibitem{foam} A. Kovner, M. Lublinsky and U. Wiedemann, JHEP {\bf 0706}, 075 (2007),
e-Print: 0705.1713 [hep-ph].





\bibitem{kovtuch} Yu. V. Kovchegov and K. Tuchin; Phys. Rev. {\bf D65}, 074026 (2002).
e-Print: hep-ph/0111362


\bibitem{Braunincl}
  M.~A.~Braun,
  Eur.\ Phys.\ J.\  C {\bf 48}, 501 (2006),
e-Print: hep-ph/0603060.

\bibitem{Baier}
  R.~Baier, A.~Kovner, M.~Nardi and U.~A.~Wiedemann,
  Phys.\ Rev.\  D {\bf 72}, 094013 (2005),
e-Print: hep-ph/0506126.
  

\bibitem{multigJK} J. Jalilian-Marian and Yu. V. Kovchegov; Phys.Rev.{\bf D70}, 114017 
(2004), Erratum-ibid. {\bf D71}, 079901 (2005);
e-Print: hep-ph/0405266. 

\bibitem{diff}
A. Kovner, M. Lublinsky and H. Weigert; Phys.Rev.{\bf D74}, 114023 (2006),
e-Print: hep-ph/0608258.

\bibitem{multig} A. Kovner and M. Lublinsky; JHEP {\bf 0611}, 083 (2006);
e-Print: hep-ph/0609227.


\bibitem{BSV}
  J.~Bartels, M.~Salvadore and G.~P.~Vacca,
  JHEP {\bf 0806}, 032 (2008)
  [arXiv:0802.2702 [hep-ph]].

\bibitem{LP}
  E.~Levin and A.~Prygarin,
  arXiv:0804.4747 [hep-ph].


\bibitem{Kovnuc}
  Y.~V.~Kovchegov,
  Nucl.\ Phys.\  A {\bf 692}, 557 (2001).
e-Print: hep-ph/0011252.





\bibitem{Balincl}
  I.~Balitsky,
  Phys.\ Rev.\  D {\bf 70}, 114030 (2004).
e-Print: hep-ph/0409314.


\bibitem{Braunincl2}
  M.~A.~Braun,
  Eur.\ Phys.\ J.\  C {\bf 55}, 377 (2008).
e-Print: 0801.0493 [hep-ph].


\bibitem{raju} 
F. Gelis, T. Lappi and R. Venugopalan, 
Phys.Rev. {\bf D78}, 054019 (2008),e-Print: arXiv:0804.2630 [hep-ph];
Phys. Rev.{\bf D78}, 054020 (2008), e-Print: arXiv:0807.1306 [hep-ph];
e-Print: arXiv:0810.4829 [hep-ph]. 


 
\bibitem{Kovpom}
  Y.~V.~Kovchegov,
  Phys.\ Rev.\  D {\bf 72}, 094009 (2005).
e-Print: hep-ph/0508276.



\bibitem{kl4} A.~Kovner and M.~Lublinsky,
Phys.\ Rev.\ D {\bf 72}, 074023 (2005).
e-Print:hep-ph/0503155.




\bibitem{yinyang} A. Kovner and M. Lublinsky;  Nucl. Phys. {\bf A779}, 220 (2006). 
e-Print Archive: hep-ph/0604085




\bibitem{herlar} A. Kovner, L. D. McLerran and H. Weigert; Phys. Rev.{\bf D52}, 6231 (1995);
e-Print: hep-ph/9502289;  
Phys. Rev.{\bf D52}, 3809 (1995);
e-Print: hep-ph/9505320.



\bibitem{mv} L.~McLerran and R.~Venugopalan, Phys. Rev. {\bf D49}, 2233 (1994); 
 e-Print: hep-ph/9309289
 Phys. Rev. {\bf D49}, 3352 (1994). 

  

  
\bibitem{BFKL}
 E. A. Kuraev, L. N. Lipatov, and F. S. Fadin,  Sov. Phys. JETP
                {\bf 45} (1977) 199 ; \\
Ya. Ya. Balitsky and L. N. Lipatov,
               {  Sov. J. Nucl. Phys.}\, {\bf 28} (1978) 22.


\bibitem{something} A. Kovner and M. Lublinsky, Phys.\ Rev.\ Lett.\  {\bf 94}, 181603 (2005), e-Print: hep-ph/0502119.

\bibitem{pertveto} C.R Schmidt, Phys. Rev. {\bf D 60}, 074003 (1999), hep-ph/9901397; \\
J.R. Forshaw, D.A. Ross and A. Sabio Vera,  Phys. Lett. {\bf B455}, 273 (1999);
e-Print: hep-ph/9903390.

\bibitem{evolveto} K. Kutak and A.M. Stasto, Eur. Phys. J.{\bf C41}, 351 (2005).
e-Print: hep-ph/0408117; 
\\
G. Chachamis , M. Lublinsky and A. Sabio Vera;  Nucl. Phys.{\bf A748},649 (2005).
e-Print: hep-ph/0408333;  \\
E. Gotsman, E. Levin, U. Maor and E. Naftali, Nucl. Phys.{\bf A750}, 391 (2005).
e-Print: hep-ph/0411242.

\bibitem{eikon}  T. Altinoluk, A. Kovner and J. Peresutti, Nucl. Phys. {\bf A}, in press;
e-Print: arXiv:0810.4533 [hep-ph].



%
%











  
  



  














 





                                  








  

              
  
          

  
  
              



  

  







 

\end{thebibliography}
\end{document}